\documentclass[3p,times]{elsarticle}

\usepackage{amsmath}

\usepackage{hyperref}
\usepackage{subfig}

\usepackage{xcolor}
\newif\ifrevision
\revisionfalse
\newcommand{\rev}[1]{\ifrevision\textcolor{blue}{#1}\else#1\fi}

\journal{Journal of Systems and Software}

\begin{document}

\begin{frontmatter}

\title{Stack Overflow Is Not Dead Yet: Crowd Answers Still Matter}


\author{Denis Helic}
\ead{dhelic@tugraz.at}
\affiliation{%
  organization={Graz University of Technology},
  city={Graz},
  country={Austria}
}

\author{Tiago Santos}
\ead{tiago.santos@e-steiermark.com}
\affiliation{%
  organization={Energie Steiermark},
  city={Graz},
  country={Austria}
}

\begin{abstract}
  Millions of users visit Stack Overflow regularly to ask community for answers to their programming questions. However, like many other platforms, Stack Overflow consistently struggles with low user retention and declining levels of user contributions to the platform. With the introduction of ChatGPT in November 2022, these ongoing difficulties on Stack Overflow were further magnified, as many users moved toward ChatGPT for programming help. In this paper, we build upon recent research on this phenomenon by analyzing the transformation of user-generated content on Stack Overflow during the post-ChatGPT period. Specifically, we analyze two years of Stack Overflow data and fit multiple causal regression models to estimate the effect of ChatGPT on the length and difficulty of user questions and code examples. We confirm an acceleration of decline in user contributions but find that ChatGPT had a significant positive effect on question and answer length, code length, and question difficulty on Stack Overflow across programming languages. Our results suggest that ChatGPT has effectively raised the bar for questions on Stack Overflow, as users increasingly turn to crowdsourced platforms for help with more complex and challenging problems. With our work we contribute to the ongoing discussion on the impact of tools such as ChatGPT on help-seeking in programming and, more broadly, on collaborative knowledge creation. Our results provide actionable insights for platform operators to support information management and user retention in the aftermath of ChatGPT's launch. 
\end{abstract}




\end{frontmatter}


\section{Introduction}
Question-and-answer (Q\&A) platforms such as Stack Overflow or similar Stack Exchange instances are used by millions of users to discuss a variety of topics. Specifically, Stack Overflow deals with programming and software development questions and attracts more than six million active users, making it the largest Web site for software developers seeking help in their daily programming tasks \citep{abdalkareem2017, rahman2018}. Despite its great success, Stack Overflow, like many other social platforms, suffers from typical problems related to online community management such as decreasing user numbers and decline in their contributions \citep{asaduzzaman2013}. Therefore, the launch of ChatGPT---a generative artificial intelligence (AI) tool---in November 2022, has immediately raised the question about the further acceleration of this trend \citep{delrio2024}. In particular, as generative AI tools demonstrate strong performance in natural language processing \citep{chang2024} and reasonably good ability to solve smaller programming problems by generating code \citep{austin2021}, our research community has quickly initiated discussions about the future relevance of Stack Overflow \citep{kabir2024}.

These initial concerns about diminishing utility of Stack Overflow in the era of ChatGPT have been reinforced by several recent research studies. For example, studies by Burtch et al. and del Rio-Chanona et al. suggest that ChatGPT is associated with an accelerated downward trend in the number of questions and answers on Stack Overflow \citep{burtch2024, delrio2024}. Moreover, two other works show that generative AI models can compete and sometimes even outperform Stack Overflow answers on several tasks including resolution of compiler errors \citep{widjojo2023} or solving privacy-related problems \citep{delile2023}. Such results suggest that ChatGPT may become a \textit{disruptive technology} with a strong potential to sustainably change software development practices including help-seeking in programming. For example, in a study of early ChatGPT adopters, Haque et al. \citep{haque2022} found a strong user expectation that ChatGPT will indeed disrupt common software development processes similarly to how ChatGPT is expected to transform the current practices in other domains such as education \citep{garcia2023}, healthcare \citep{varghese2024}, or scientific publishing \citep{gao2023}.

In this manuscript, we aim to shed light on the transformative potential of ChatGPT on Stack Overflow. In particular, we build upon previous studies on accelerated decline in contributions to Stack Overflow \citep{burtch2024, delrio2024} by asking the question about the effect of ChatGPT on the content of those contributions. 
Specifically, we ask whether the introduction of ChatGPT has a profound effect on the difficulty levels of the user questions post-ChatGPT. We hypothesize that the Stack Overflow users may turn to ChatGPT more frequently for answers to smaller-scale and elementary programming problems, but that the Stack Overflow community still might play a major role in finding answers to larger-scale and advanced programming questions. To answer our research question, we analyze two years of Stack Overflow data (around six millions of user posts), including all questions and answers six months before the introduction of ChatGPT as well as six months after that event. In addition, we also include the data from the same time period one year before as a control group. Then, we extract a variety of features such as question length or the length of code examples from our collected data. Moreover, using a pretrained large-scale neural embedding model for programming questions and code as well as a ground truth dataset with three difficulty levels for programming questions, we first embed all Stack Overflow questions and then construct a machine learning model to predict the difficulty level of those questions. To control for thematic categories of questions, we divide our data according to the user-created tags into question related to \textit{python}, \textit{java}, or \textit{web} development and repeat all the calculations for each of those groups. Lastly, to estimate the effect of ChatGPT, as well as the short-term and long-terms trends in ChatGPT effect on our content features, we fit a series of causal regression models while controlling for the thematic group and general temporal trends.

While we confirm the previous results on the shrinking volume of user contributions to Stack Overflow, we find that ChatGPT has a differential effect on the content of those contributions. Specifically, we find a significant positive effect of ChatGPT on the question and answer length, code length, as well as the question difficulty. For example, expressed as a percentage of the standard deviation of the given quantity, we observe a $6\%$ increase in question length and a $5\%$ increase in answer length, six months after the ChatGPT launch. In addition, we find a strong, consistent, and positive trend of the ChatGPT effect on those quantities over the whole period, i.e., in short-term (a few weeks after the launch) as well as long-term periods (a few months after the launch). Moreover, our results are robust to disaggregation of the questions according to the user tags. For instance, we find a $21\%$ increase in length of \textit{python} code examples or an $11\%$ increase in \textit{java} question difficulty at the scale of individual standard deviations at the end of our observation period. However, we find that the effects are stronger and more stable for tags with larger numbers of questions such as \textit{python} or \textit{web} development. \rev{Additionally, we also observe heterogeneous effects across user segments, with regular users exhibiting the strongest response.} Our large-scale results suggest that Stack Overflow community experiences a sustained behavioral change in the aftermath of the ChatGPT start. We argue that due to ChatGPT, users ask the Stack Overflow crowd fewer of elementary and more of advanced programming questions. While there are less questions to Stack Overflow in total, these questions shift towards higher difficulty levels, and the crowd answers to those questions are still important---Stack Overflow is far from being dead, evolving further as a consequence of the new technological advancements such as ChatGPT.

With our work we contribute to the ongoing discussion on the effect of generative AI tools, here ChatGPT, on software development practices and help-seeking in programming. Going beyond software development, we also frame our work within the broader discussion on the consequences of modern AI on collaborative knowledge creation, as such advanced tools, capable of generating text and code, could change practices on other knowledge platforms not limited to Q\&A platforms such as Wikipedia. Our results provide actionable information for platform operators that can help in information management and organization, as well as in onboarding and retaining users. Moreover, we also give initial insights into the drifts of the user content post-ChatGPT that may be helpful for training and education in programming, as well as for training new versions of such AI models. \rev{Finally, to ensure reproducibility and verifiability of our results, we publish all of our code.\footnote{\url{https://github.com/dhelic/so}}}

\section{Related work}


\subsection{\rev{ChatGPT and software development}}
Technically, ChatGPT belongs to a class of large language models (LLM), a particular type of generative AI tools. LLMs comprise billions of parameters in deep neural network architectures that are trained and fine-tuned on huge textual datasets. Once trained, users interact with the LLMs by writing prompts, and LLMs generate corresponding answers in response to those prompts. For a more in-depth information and an overview of architectures, models, training, fine-tuning, and prompting strategies for a wide range of LLMs, we refer an \rev{interested} reader to the paper by Naveed et al. ~\citep{naveed2025}.

Going beyond natural language processing \citep{chang2024}, ChatGPT also demonstrated its remarkable capabilities on various tasks in diverse domains \citep{teubner2023} including education \citep{garcia2023}, medicine \citep{chow2023, heng2023}, or even research \citep{gao2023}. Moreover, in the area of software development, LLMs such as ChatGPT and Copilot have received a lot of attention from practitioners and researchers, in particular for the task of code generation \citep{liu2023}. For example, LLMs demonstrated outstanding code generation performance on popular code completion benchmarks \citep{austin2021, chen2021}, even further improved by instruction fine-tuning \citep{luo2023}. Similarly, recent research also demonstrated the ability of LLMs in locating bugs \citep{jesse2023} or resolving simple issues from \rev{GitHub} \citep{jimenez2024}. While these initial results potentially suggest a range of useful applications of LLMs in programming practice, several other studies caution about potential issues related to such applications. For example, Siddiq et al. \citep{siddiq2022} criticize poor quality of the generated code that frequently includes too long methods or code duplicates. Regarding code security, Perry et al. \citep{perry2023} found that the code generated by AI is typically less secure, while the work by Sandoval et al. \citep{sandoval2023} suggested that code written in collaboration between programmers and LLMs is generally less secure, but the difference to the control group (programmers only) is not substantial. In another line of research, Vaithilingam et al. \citep{vaithilingam2022} evaluated the usability of LLMs for code generation and found that the LLM frequently provided useful starting points and saved time for searching information online, but at the same the tool usage was associated with understanding difficulties and reduced task-solving efficiency.

In this paper, we analyze another aspect of software development that may have been affected by LLMs, that of collaborative knowledge creation and sharing in software domain. Hence, we ask the question whether such advanced tools, capable of generating code and answering programming questions, have changed online community practices on large software-related Q\&A platforms. For example, users may increasingly turn to ChatGPT as a starting information seeking activity \citep{vaithilingam2022}, which may lead to a decline in knowledge sharing on the platforms as the first evidence suggests \citep{delrio2024}. Nevertheless, while ChatGPT may support users in simple or programming tasks easily solved with a quick lookup in vast software knowledge databases, we aim to investigate whether users still may seek programming help from the community in case of more complex coding problems.

\subsection{\rev{Software knowledge platforms and ChatGPT}} 
Currently, Stack Overflow is the most successful help-seeking online community, used by a large number of programmers to ask questions on various programming topics \citep{abdalkareem2017, rahman2018}. Several studies investigated the reasons for Stack Overflow's popularity suggesting quick response times \citep{mamykina2011}, high quality answers to conceptual questions \citep{treude2011}, potential for usable code snippets \citep{yang2016}, or possibility to ask question from a broad range of topics and types \citep{allamanis2013}, as potential factors contributing to the Stack Overflow success. Nevertheless, Stack Overflow recently experienced decreasing numbers of users, questions, and answers \citep{delrio2024} and the community started analyzing the potential reasons for this development. For example, Stack Overflow community has been recognized as a community that is unwelcoming to the newcomers,\footnote{\url{https://stackoverflow.blog/2018/04/26/stack-overflow-isnt-very-welcoming-its-time-for-that-to-change}, last retrieved on September 06, 2025.} resulting in problems of attracting new users. As the online communities need to have a healthy mix of experienced, expert, and novice users for sustainable activity levels \citep{santos2019, santos2019a}, this caused a set of measures by the Stack Overflow operators to promote more welcoming culture including, for instance, a new user badge \citep{santos2020}. Also, toxicity and negative sentiment of a substantial amount of answers on Stack Overflow \citep{asaduzzaman2013} was found to be related to the problem of retaining less experienced users. Moreover, Calefato et al. found that to enhance their chances of becoming a useful answer users need to carefully frame and pose their questions  \citep{calefato2018}.

From the technical perspective, the quality of code examples from Stack Overflow has been also analyzed in past research. For example, Yang et al. \citep{yang2016} found that the majority of code examples in answers can not be compiled. Similarly, Zhong et al. \citep{zhong2024} analyzed the API use in the code examples from Stack Overflow answers and found that a substantial fraction of these examples violated the rules on how to write API calls. Moreover, the study by Fischer et al. \citep{fischer2017} investigated the security risks from the Stack Overflow code examples and found a substantial amount of examples containing insecure code snippets. While these studies suggest caution when using Stack Overflow and the code examples from the community, the value of this collaborative knowledge platform \citep{storey2010} as a support tool for software development \citep{vasilescu2013} is widely recognized. \rev{For example, Stack Overflow has been successfully integrated in software developer environments in the form of plugins \cite{ponzanelli2013, ponzanelli2014}, allowing developers to prompt Stack Overflows's crowd knowledge with structured queries and include search results directly into their current programming projects. Along those lines, researchers have also evaluated automatically generated natural language queries for searching in Stack Overflow (similar to LLM prompts) \cite{rahman2016}, queries augmented with the current programming context (similar to LLM context creation) \cite{rubei2020}, or automatically generated paraphrased questions for querying Stack Overflow \cite{gao2023a}.} Similarly, de Souza et al. \citep{desouza2014} presented an approach based on the ``crowd knowledge'' from Stack Overflow to recommend information to the developers supporting their current activities. Additionally, the researchers have used the vast Stack Overflow knowledge database to extract code and corresponding textual descriptions \citep{yin2018}, or to summarize source code for future references \citep{iyer2016, kou2023}. 

Apart from already mentioned studies on the interplay between Stack Overflow and ChatGPT \citep{burtch2024, delile2023, delrio2024, kabir2024, widjojo2023}, several other studies focused on a particular use case of LLMs as developer's assistant, a role typically occupied by platforms such as Stack Overflow. These studies found empirical evidence for an increased developer productivity while using LLMs \citep{peng2023, ross2023}, the ability of LLMs to act as pair programmers for expert developers \citep{moradidakhel2023}, or acceptable code quality \citep{xu2022, yetistiren2022}. Such capabilities combined with the immediate response, infinite repeatability, and possibility to interactively refine questions could induce dramatic drops in the user-contributed content on software Q\&A platforms, effectively also reducing the training data for future LLMs. This situation caused a prompt action (within the first week since the ChatGPT launch) from the Stack Overflow operators, banning LLMs when posting content.\footnote{\url{https://meta.stackoverflow.com/questions/421831/policy-generative-ai-e-g-chatgpt-is-banned}, last retrieved on September 06, 2025.}

In this paper, we build upon those studies by further investigating the interplay between ChatGPT and collaborative knowledge platforms in the domain of software development. While previous studies identified substantial changes in Stack Overflow following the ChatGPT launch such as decline in the contributed content \citep{burtch2024, delrio2024} or increase in text complexity as measured by the word lengths \citep{burtch2024}, in our study we also analyze the contributed code examples and ask the question whether the community shifted towards more difficult questions due to the availability of LLMs.

\subsection{\rev{ChatGPT as a disruptive technology}}
Since the launch of ChatGPT in November of 2022, there was an active public and scientific discussion of its potential disruptive consequences on knowledge creation, education, software development, medicine, or entertainment industry. For example, Garcia \citep{garcia2023} discussed the potential improvements or disruption of educational practices through ChatGPT concluding that ChatGPT will induce enduring changes in education by requiring from educators to learn and adopt this new technology in their teaching practice. Moreover, introduction of ChatGPT generated a lot of discussion in the academic community on the ability of ChatGPT and similar technologies to write scientific papers and abstracts. For example, Gao et al. \citep{gao2023} compared the ChatGPT generated abstracts with the genuine scientific abstracts and found that both automatic and human detectors can detect a large proportion of ChatGPT generated abstracts albeit not all of them. Hence, the authors concluded that journals and publishers will need to adopt new editorial and reviewing policies to deal with this new situation. In other scientific domains, such as medicine or healthcare, researchers investigated and discussed the effects of ChatGPT on medical education \citep{heng2023}, medical chatbots \citep{chow2023}, or the transformative influence on healthcare \citep{varghese2024}. These and similar studies draw similar conclusions: while ChatGPT and similar LLM technologies have a potential as a disruptive technology to improve processes, practices, and the quality of interactions, there are also concerns related to the quality, correctness, transparency, fairness and ethical issues related to a wide adoption of this new technology \citep{guo2023, shen2023, ye2023}.

The discussion of ChatGPT and LLMs as a disruptive technology, taps into a broader economic theory of disruptive technology and disruptive innovation, introduced by Christensen in his 1997 book ``The innovator's dilemma: when new technologies cause great firms to fail'' \citep{christensen1997}. The original theory defined a disruptive technology as a technology that is inferior in the main attributes of the mainstream technology but concentrates on alternative attributes that the mainstream technology neglects. Nevertheless, as the new technology matures, it surpasses the dominant technology, first in specific markets, and then potentially also in other more general markets as well. The theory of disruptive technology was later extended to the concept of a disruptive innovation, which is not only focused on technology but may include, among others, disruption in business models or products \citep{christensen2003, hang2015, markides2006}. A comprehensive survey on the economic aspects of the theory of disruptive technology and innovation can be found in \citep{si2020}. 

In this paper, we concentrate on the impact and the (disruptive) change that introduction of ChatGPT can potentially have on online knowledge creation communities, in particular, on collaborative software Q\&A platform Stack Overflow. While prior work \citep{burtch2024, delrio2024} has already identified a negative quantitative impact of ChatGPT on user contributions on Stack Overflow, we focus on the content related aspects of user posts such as questions difficulty and the complexity of included code examples in the post-ChatGPT period.

\section{Descriptive analysis}

\subsection{\rev{Dataset}}
We study Stack Overflow, the largest question and answering (Q\&A) community for seeking help with programming and software development problems. We obtain the data from the official data dumps from September 2008 until March 2024.\footnote{\url{https://archive.org/details/stackexchange}, last retrieved on September 06, 2025.} From these data dumps, we extract all posts from a two years period starting on May 31, 2021 until May 28, 2023. We split this data in two parts: (i) 21/22 period starting May 31, 2021 until May 29, 2022 and (ii) 22/23 period from May 30, 2022 until May 28, 2023. Hence, 22/23 period includes six months of data prior to the ChatGPT launch on November 30, 2022 and six months of data after the launch. 21/22 period contains the same time interval one year prior to the ChatGPT launch and serves as the control data for our analysis. 

The extracted data contains slightly more than six million ($6,000,198$) posts with $2,642,840$ questions and $3,352,591$ answers. The remaining $4,767$ posts are related to internal communication and administration of the system. The data contains time of the post, title, the content of the post included as HTML code, username, last edit time, score, view count, comment count, the accepted answer and tags (if the post is a question), and several other administrative fields. We start our analysis by parsing the HTML code to extract the textual content as well as the content of \textit{$<$code$>$} elements that include verbatim code examples that users post in their questions and answers. After removing posts with invalid HTML code ($2,435$ posts) we have slightly less than six million posts ($5,997,763$). Further, after selecting only question and answer posts by eliminating administrative posts our final dataset includes $2,642,730$ questions and $3,352,332$ answers. In 21/22 period we have $3,283,819$ posts comprising $1,438,850$ questions and $1,843,515$ answers. In 22/23 period we obtain approximately half a million posts less ($2,713,944$) including $1,203,880$ questions and $1,508,817$ answers. We summarize these basic dataset statistics in Table \ref{tab:dataset}.

\begin{table}[t]
\centering
\caption{\textit{Dataset statistics.} We show the number of posts including questions and answers in our two years observation period before and after parsing the HTML content of the posts. We observe a downwards trend in the number of posts in the second year.}
\begin{tabular}{l|r|r|r|r}\hline
 & \textbf{Period} & \textbf{Total} & \textbf{Questions} & \textbf{Answers}\\\hline\hline
Initial dataset & Both & $6,000,198$ & $2,642,840$ &  $3,352,591$\\\hline\hline
Dataset after parsing HTML & Both & $5,997,763$ & $2,642,730$ & $3,352,332$\\
& 21/22 & $3,283,819$ & $1,438,850$ & $1,843,515$\\
& 22/23 & $2,713,944$ & $1,203,880$ & $1,508,817$\\\hline 
\end{tabular}
\label{tab:dataset}
\end{table}

\begin{figure}[t!]
	\centering
	\subfloat[Question Volume]{\includegraphics[width=0.49\textwidth]{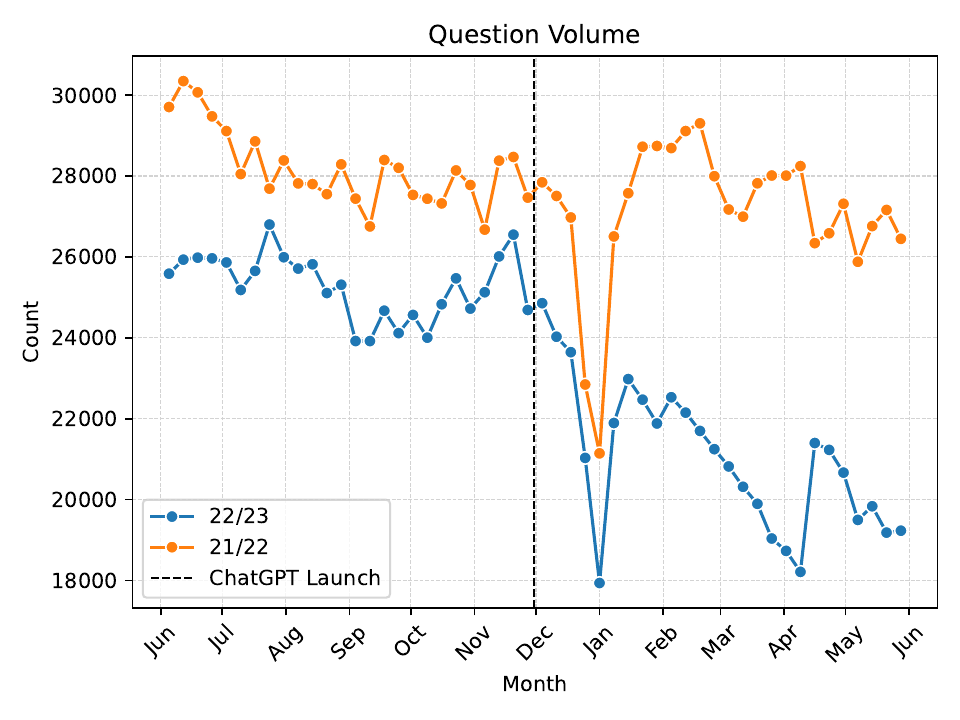}\label{fig:q}}
	\subfloat[Question views]{\includegraphics[width=0.49\textwidth]{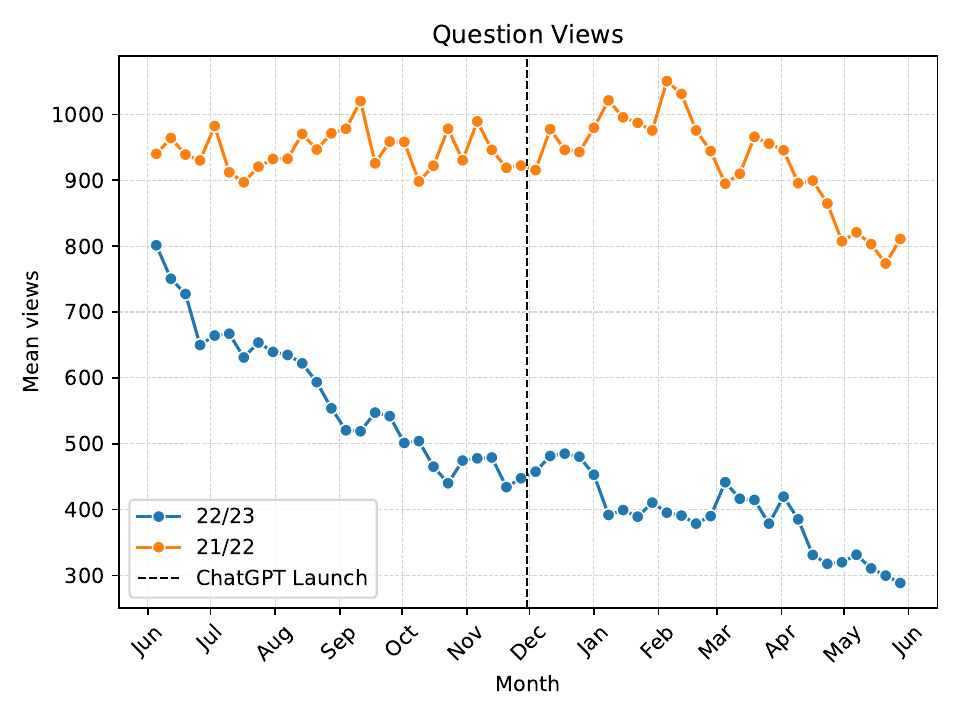}\label{fig:q_views}}\\
    \subfloat[Question score]{\includegraphics[width=0.49\textwidth]{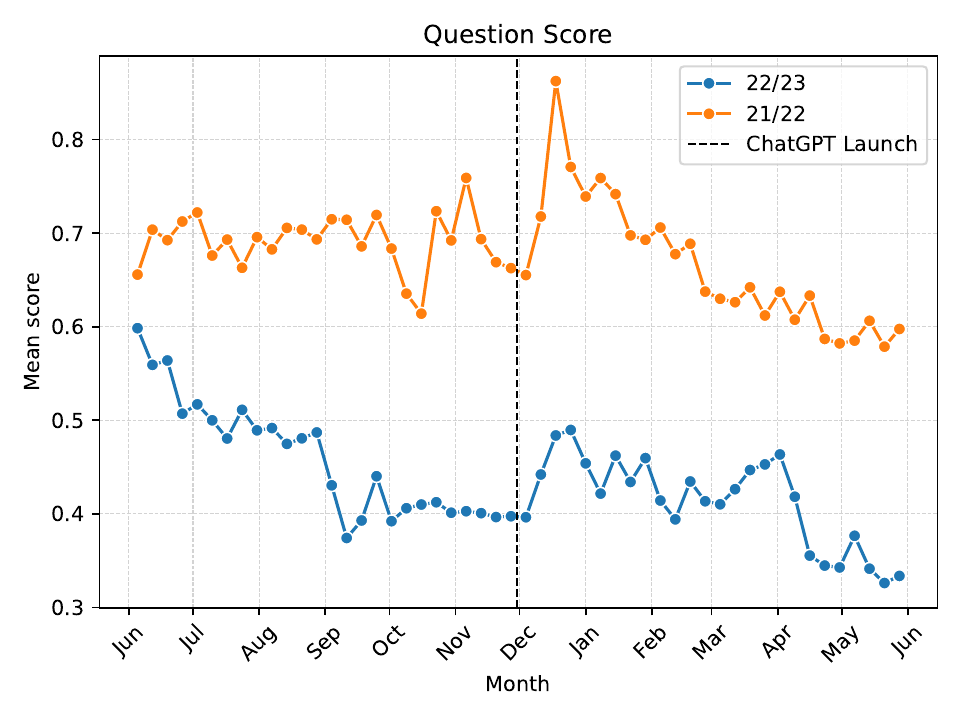}\label{fig:q_score}}
    \subfloat[Question length]{\includegraphics[width=0.49\textwidth]{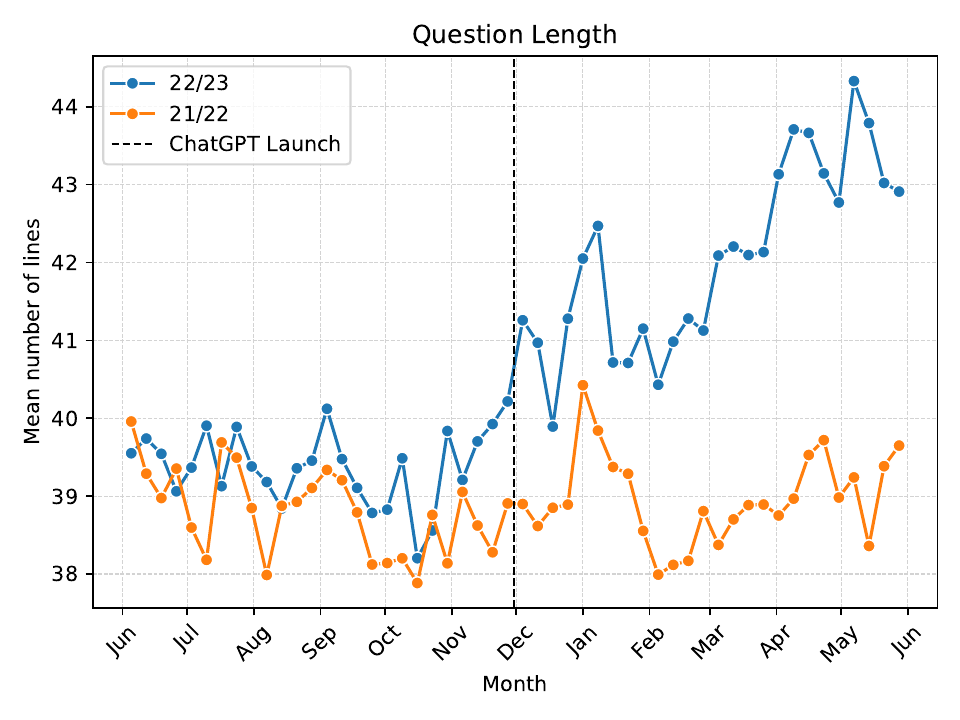}\label{fig:q_line_counts}}
	\caption{\textbf{Questions on Stack Overflow.} We compare the volume (a), views (b), score (c), and number of lines (d) in questions on Stack Overflow in 22/23 (May 30, 2022 through May 28, 2023) and 21/22 periods (May 31, 2021 through May 29, 2022). The launch of ChatGPT is in the middle of the 22/23 period (November 30, 2022). Stack Overflow already experienced an ongoing downwards trend in the number of questions, question views, and question scores, even prior to ChatGPT (cf. orange lines, as well as blue lines before the ChatGPT launch in (a), (b), and (c)). Similarly to previous work \citep{burtch2024, delrio2024}, we also observe an accelerating negative trend in (a) after the ChatGPT launch (cf. slope of the blue and orange lines post-ChatGPT). We do not observe such an acceleration in the question views in (b) and question scores in (c) but more of a continuing negative trend that already started in the six months period before ChatGPT. However, when comparing the length of the questions between the 22/23 and 21/22 periods, we see a substantial change around the ChatGPT launch in (d). Particularly, while throughout the whole 21/22 period the number of lines in questions is rather stable (orange line), we observe a strong upwards trend in the 22/23 period after the ChatGPT launch (positive slope of the blue line), suggesting a substantial and sustained increase in the question length in that period.}
	\label{fig:questions}
\end{figure}

\subsection{\rev{Post volume and length}}
\begin{figure}[t]
	\centering
	\subfloat[Answer Volume]{\includegraphics[width=0.33\textwidth]{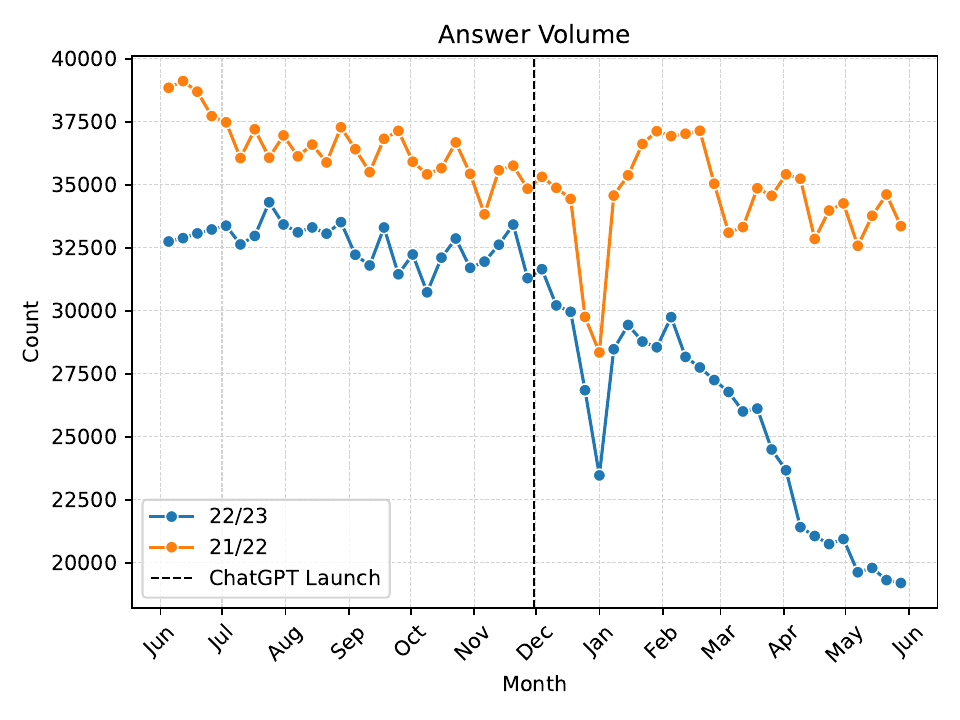}\label{fig:a}}
    \subfloat[Answer score]{\includegraphics[width=0.33\textwidth]{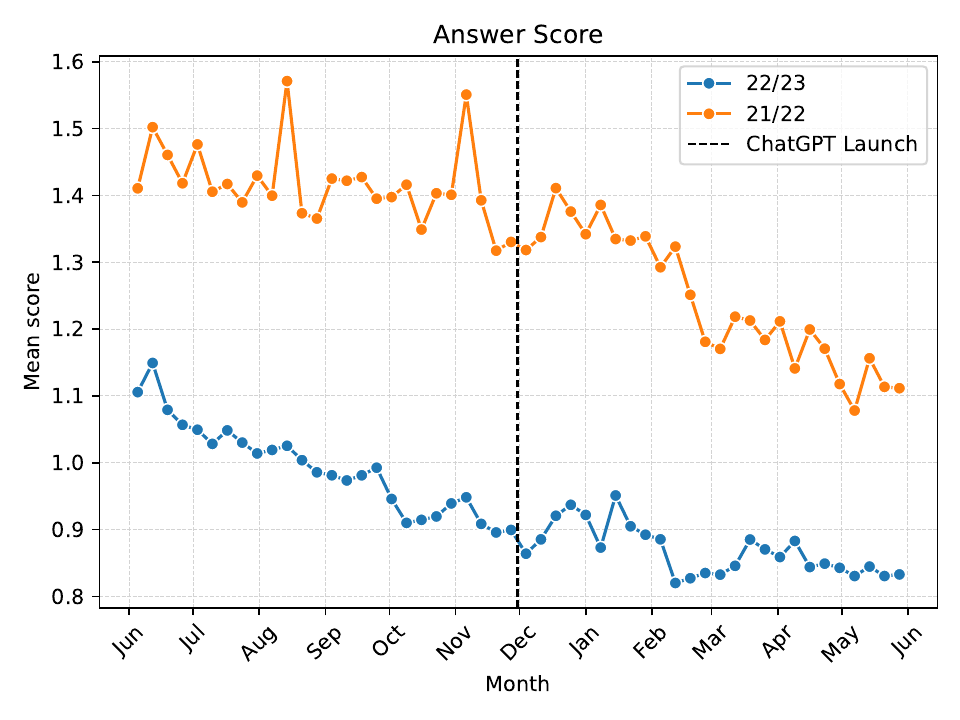}\label{fig:a_score}}
    \subfloat[Answer length]{\includegraphics[width=0.33\textwidth]{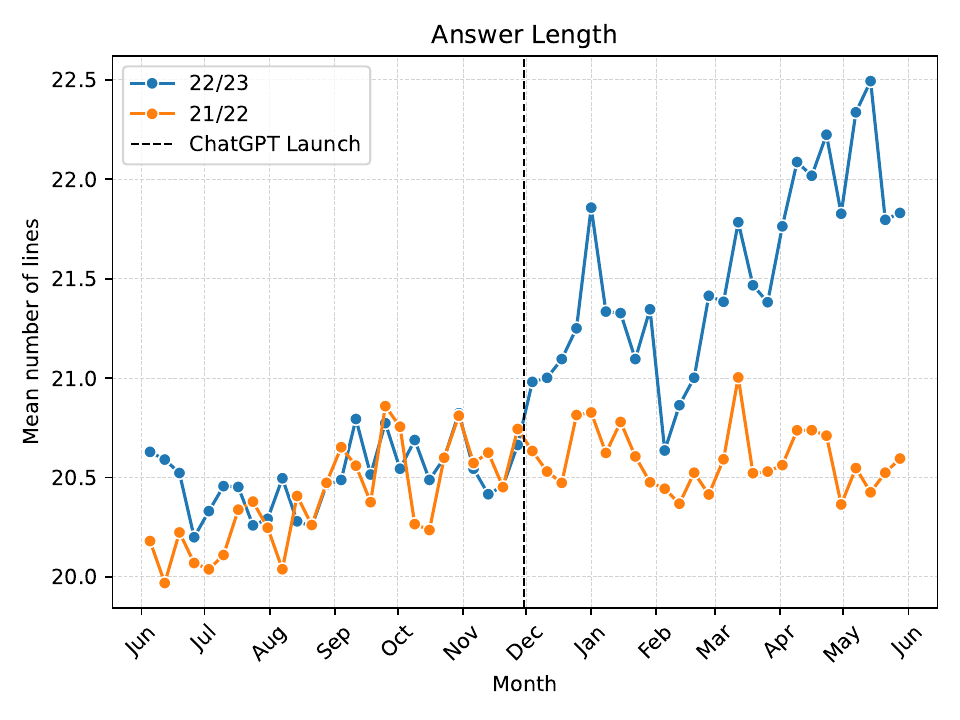}\label{fig:a_line_counts}}
	\caption{\textbf{Answers on Stack Overflow.} We compare the volume (a), score (b), and number of lines (c) in answers on Stack Overflow between 22/23 period and 21/22 periods. We observe a similar temporal evolution as in questions (cf. Figure \ref{fig:questions}). In particular, the negative trend in the answer volume accelerates after the ChatGPT start (cf. blue line slope in (a)), while there is no visible change of an already existing negative trend in the answer score in (b). Finally, in (c) we observe an accelerating positive trend in the length of answers after the ChatGPT start.}
	\label{fig:answers}
\end{figure}

We start by comparing basic weekly statistics of posts between periods 22/23 and 21/22. Period 22/23 is treated with the ChatGPT launch while period 21/22 serves as a control period. In Figure \ref{fig:questions}, we show the weekly question volume (number of questions posted), and weekly means\footnote{\rev{In addition, we also investigate weekly medians of these variables, which we do not show, as we observe qualitatively similar results.}} of question views, question scores and question length, which we compute by counting the number of lines in the post's HTML code.\footnote{\rev{Note that in all such figures, we do not start y-axis at zero for presentation purposes. This makes interpretation of the results more difficult as the magnitude of the trends may be either amplified or diminished by this particular choice of visualization. However, to avoid misinterpretation we accompany these visualizations with exact numerical characterization of the trends in the main text.}} In Figure \ref{fig:answers}, we show the same statistics for answers, except for answer views, which are not recorded separately but are subsumed in the views of the corresponding questions. For both post types we reproduce the results from previous studies \citep{burtch2024, delrio2024} finding a sharp decline in the post volume since the launch of ChatGPT (cf. slopes of blue lines in Figures \ref{fig:q} and \ref{fig:a}). This decline accelerates an already existing negative trend (cf. slope of orange lines in Figures \ref{fig:q} and \ref{fig:a}) in post volume on Stack Overflow suggesting an essential shift in user posting behavior post-ChatGPT. Moreover, the trend is visible over a long time span and persists until the end of our observation period in May 2023, six months after the ChatGPT launch. 

Opposite to question volume, we do not observe an accelerating downwards trend in the mean question views post-ChatGPT (see Figure \ref{fig:q_views}). In particular, we observe a continuing and strong negative trend that already started prior to the ChatGPT launch (cf. blue line in Figure \ref{fig:q_views}). This negative trend results in a considerable mean question view drop in the last weeks of 22/23 period (around 300 weekly mean views) as compared to the same time one year before (more than 800 weekly mean views), a drop of around $60\%$ of the last year's weekly average views. We observe a similar behavior in question and answer scores in Figures \ref{fig:q_score} and \ref{fig:a_score}---an already existing downwards trend in both question and answer scores continues until the end of the observation period without a visible acceleration post-ChatGPT. This trend results in approximately $50\%$ drop in question scores and  $40\%$ drop in the answer scores between the beginning and the end of our observation period.

However, unlike post volume and scores, or question views, the length of both questions and answers undergoes a major and enduring change after the ChatGPT introduction, see Figure \ref{fig:q_line_counts} and Figure \ref{fig:a_line_counts}. While both question and answer length exhibit a stable oscillating behavior (around $39$ lines for questions, and $20.5$ lines for answers) without any visible trends prior to the ChatGPT, both quantities demonstrate a strong, robust, and ongoing upwards trend in six months after the ChatGPT start. Mean question length at the end of the observation period is around $43.5$ lines ($11.5\%$ increase) while mean answer length is approximately $22$ lines ($7\%$ increase). This extensive change in post's length combined with dwindling post volume suggests a fundamental behavioral trend on Stack Overflow---while users do not seek help on Stack Overflow as often as before, when they do, they tend to write longer questions.

\subsection{\rev{Tags}}
On Stack Overflow users can assign up to five unique tags per question for categorization purposes. The tags come from a predefined vocabulary, which more experienced users with enough reputation can extend by defining new tags.\footnote{\url{https://stackoverflow.com/help/privileges/create-tags}, last retrieved on September 06, 2025.} To account for possibility of aggregation bias when analyzing heterogeneous data such as Stack Overflow questions, we divide the questions into groups according to their tags. In particular, we are interested in analyzing whether an association observed in aggregated data (in our case the association between the ChatGPT launch and the volume or length of the questions) disappears or reverses when data is divided into the underlying groups \citep{mehrabi2022} (here, defined by the question tags), a paradox known as Simpson's paradox \citep{blyth1972}.

\begin{figure}[t!]
	\centering
    \subfloat[Top 10 tags overall]{\includegraphics[width=0.33\textwidth]{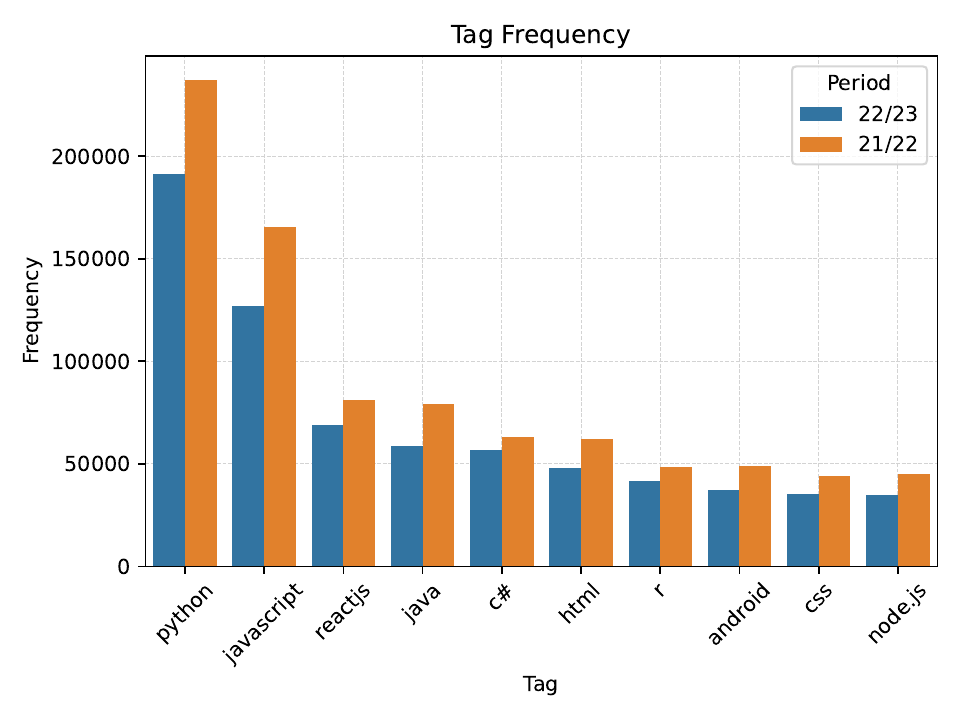}\label{fig:ttc}}
	\subfloat[Top 10 tags in 22/23]{\includegraphics[width=0.33\textwidth]{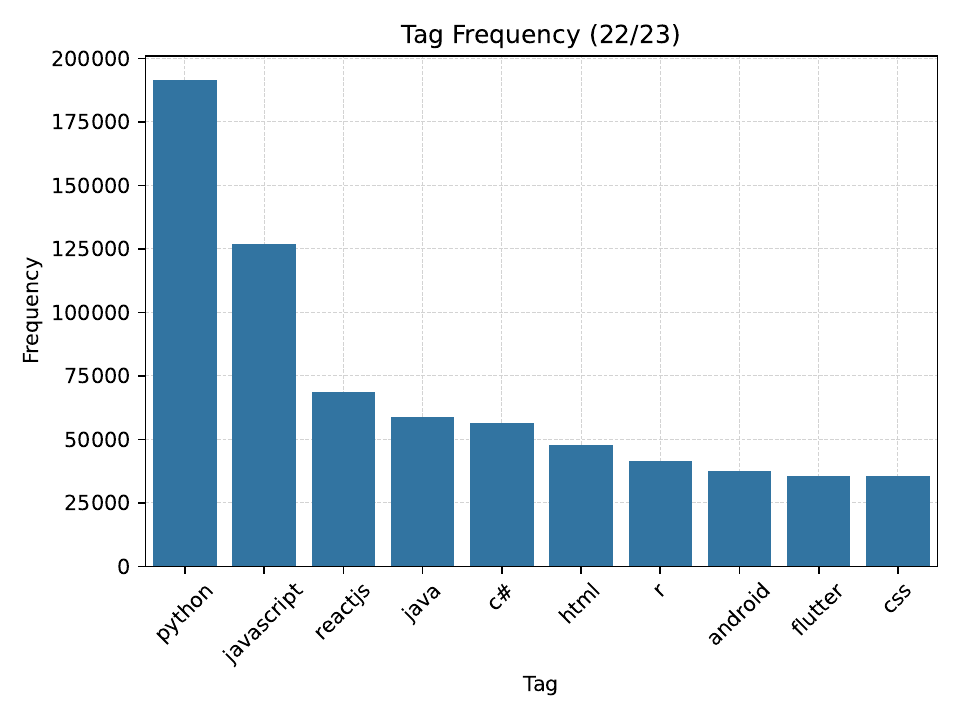}\label{fig:tt}}
    \subfloat[Top 10 tags in 21/22]{\includegraphics[width=0.33\textwidth]{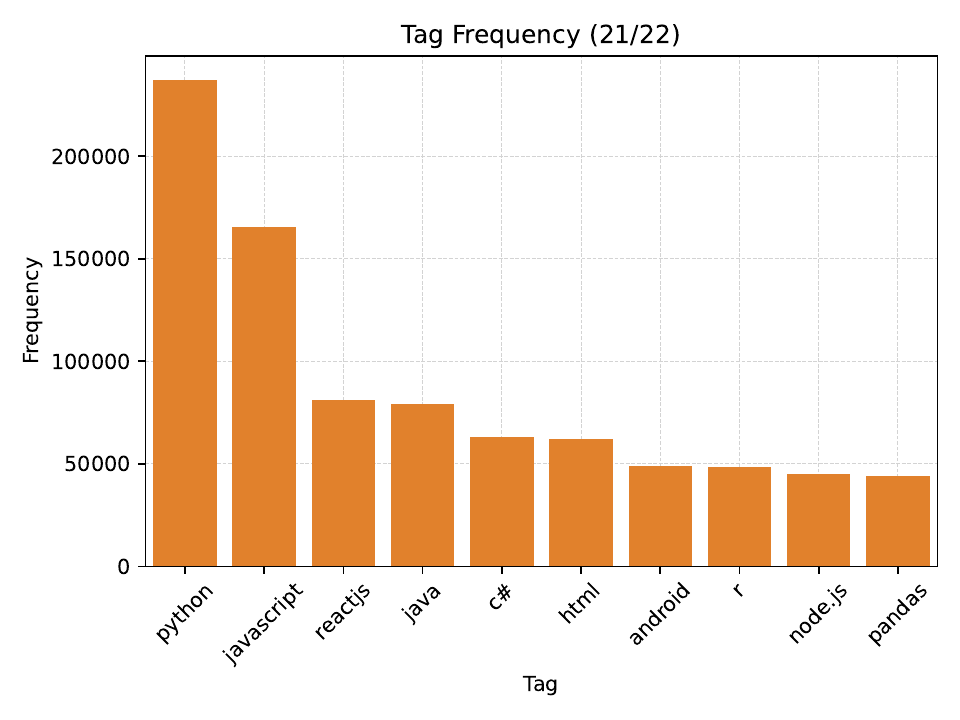}\label{fig:tc}}
	\caption{\textbf{Tags on Stack Overflow.} In (a) we show top 10 tags on Stack Overflow in our two-year observation period and compare their volume for periods 22/23 and 21/22. We observe a substantially lower tag volume in the period 22/23 as compared to 21/22, similar to the decreasing question and answer volume. In (b) and (c) we show top 10 tags for both periods individually. While the top seven tags including e.g., \textit{python}, \textit{javascript}, or \textit{reactjs} remain unchanged, there are some fluctuations among remaining tags with \textit{node.js} and \textit{pandas} dropping out of top 10 tags in 22/23.}
	\label{fig:tags}
\end{figure}

\begin{figure}[t!]
	\centering
	\subfloat[Web: question volume]{\includegraphics[width=0.33\textwidth]{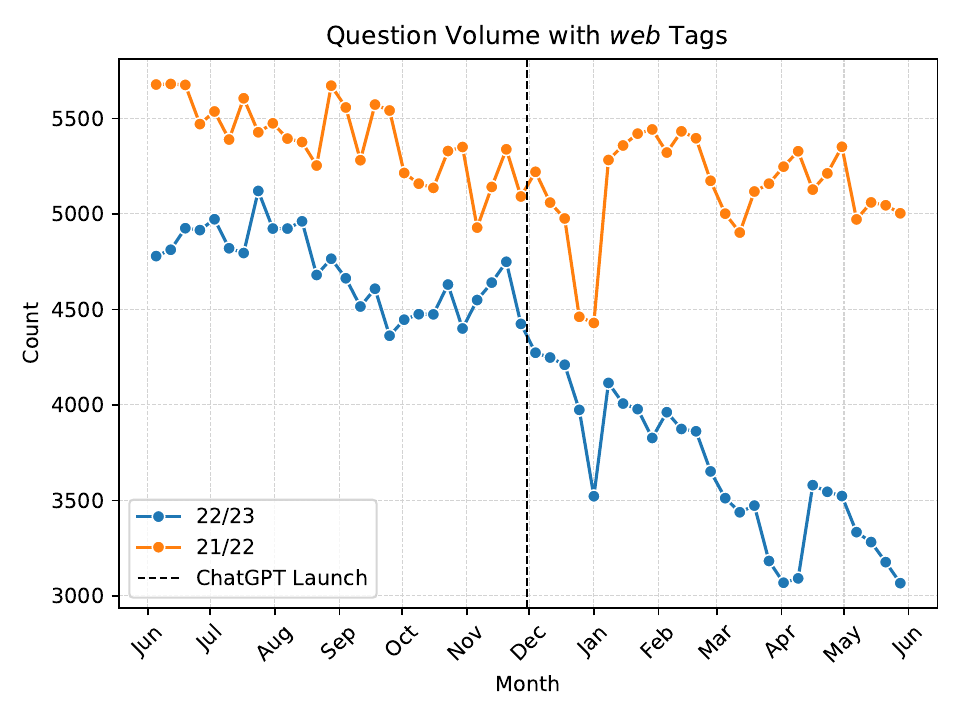}\label{fig:t_web}}
	\subfloat[Web: question length]{\includegraphics[width=0.33\textwidth]{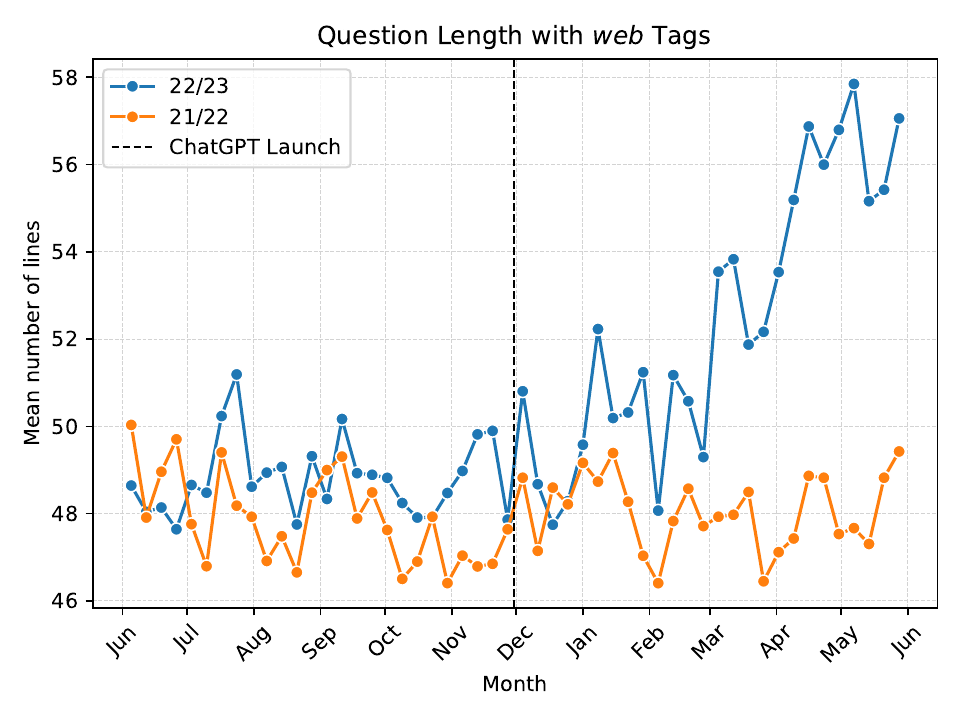}\label{fig:t_web_lc}}
    \subfloat[Web: lines of code]{\includegraphics[width=0.33\textwidth]{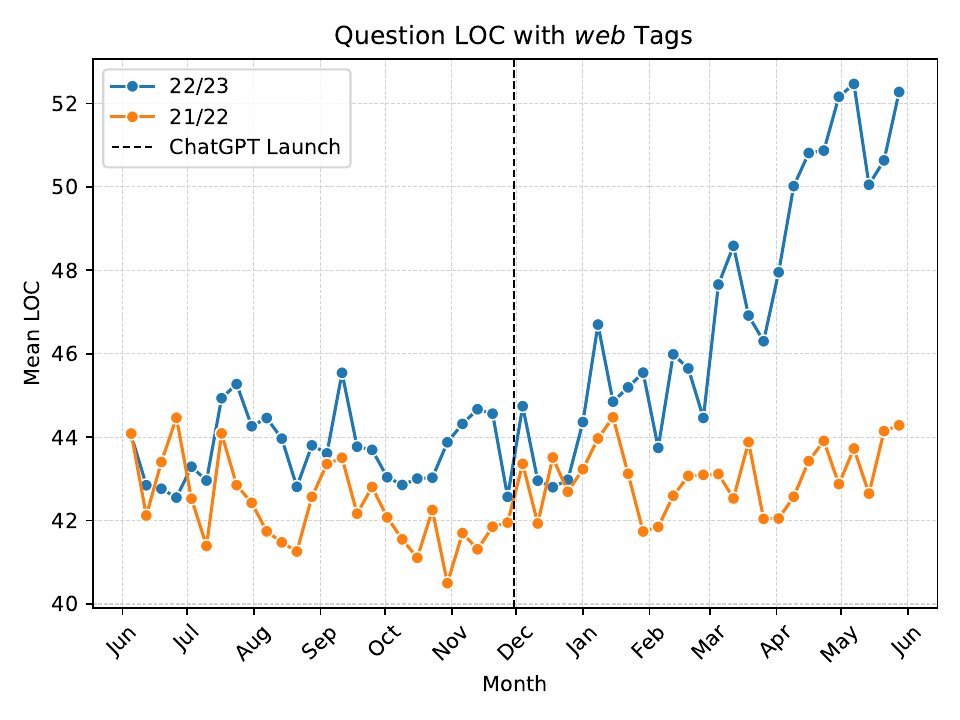}\label{fig:t_web_loc}}\\
    \subfloat[Python: question volume]{\includegraphics[width=0.33\textwidth]{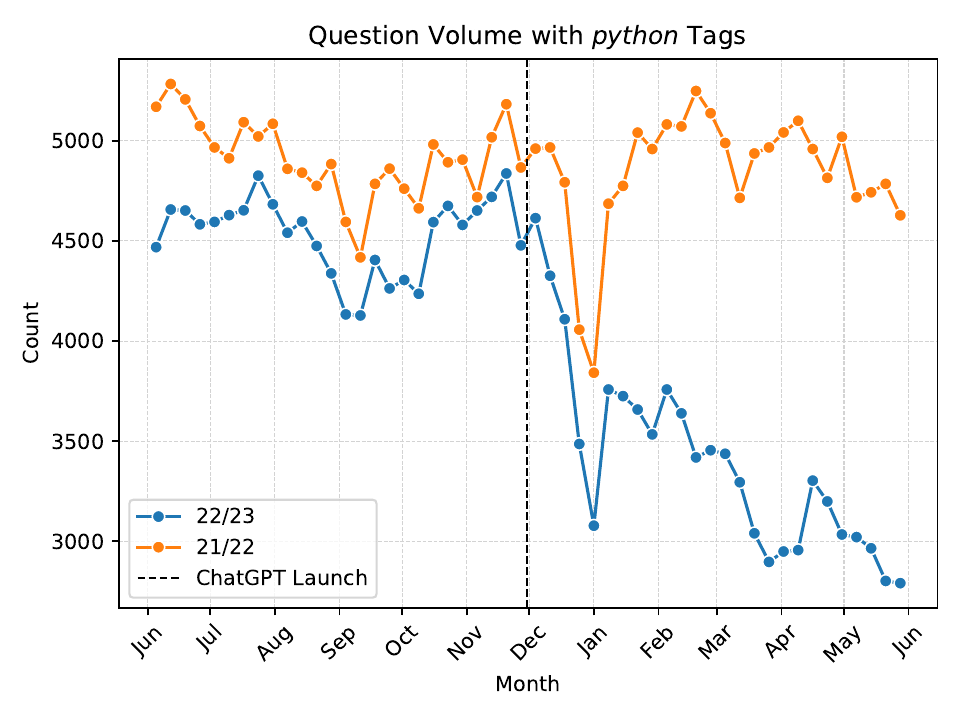}\label{fig:t_python}}
	\subfloat[Python: question length]{\includegraphics[width=0.33\textwidth]{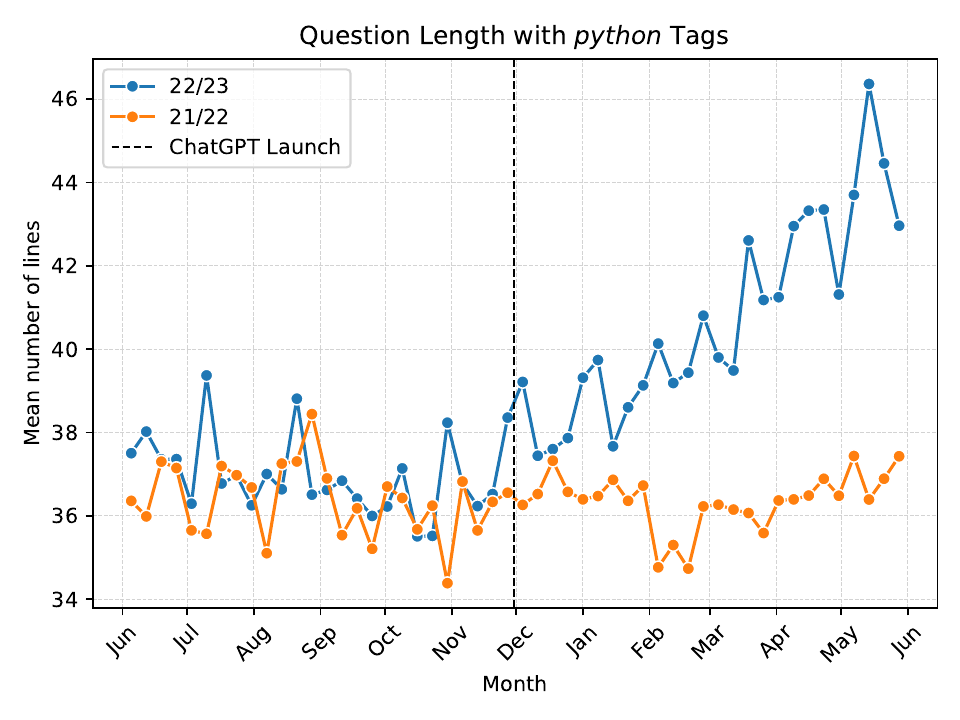}\label{fig:t_python_lc}}
    \subfloat[Python: lines of code]{\includegraphics[width=0.33\textwidth]{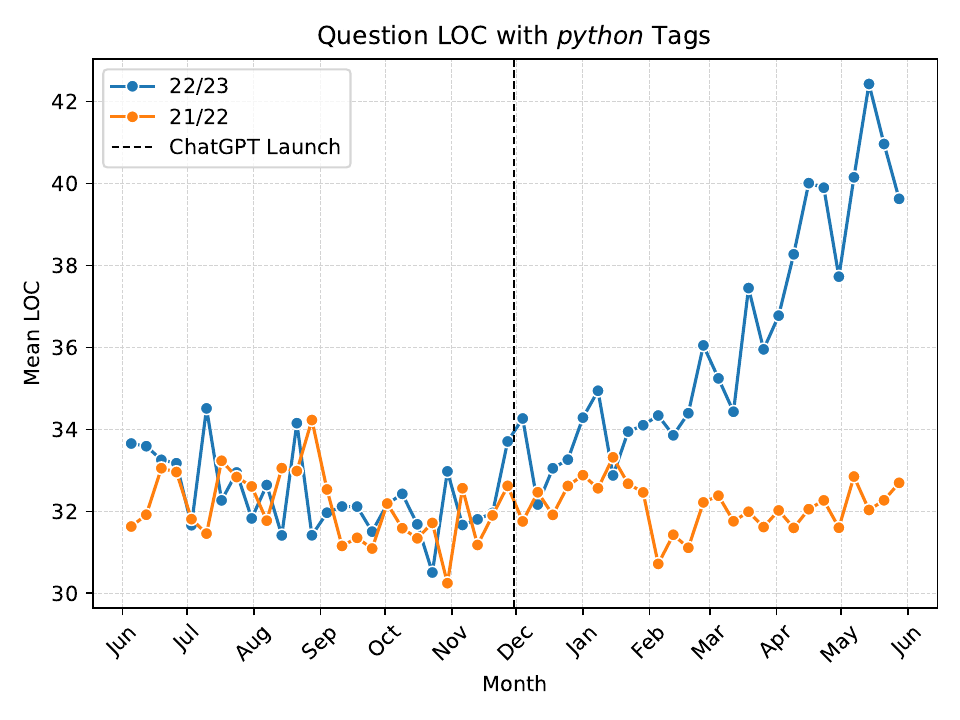}\label{fig:t_python_loc}}\\
    \subfloat[Java: question volume]{\includegraphics[width=0.33\textwidth]{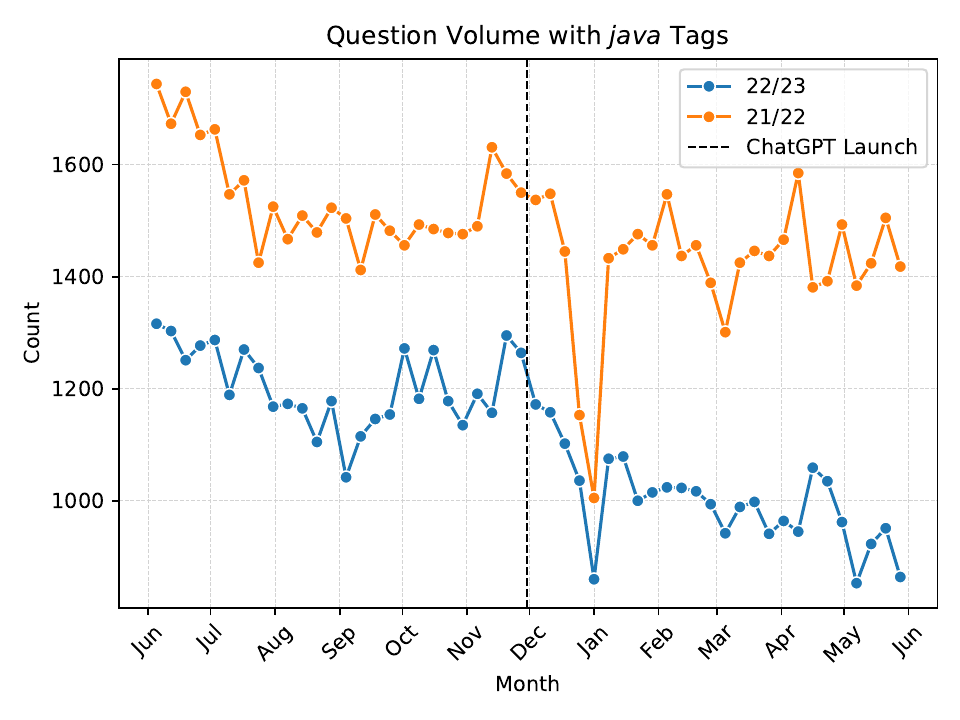}\label{fig:t_java}}
	\subfloat[Java: question length]{\includegraphics[width=0.33\textwidth]{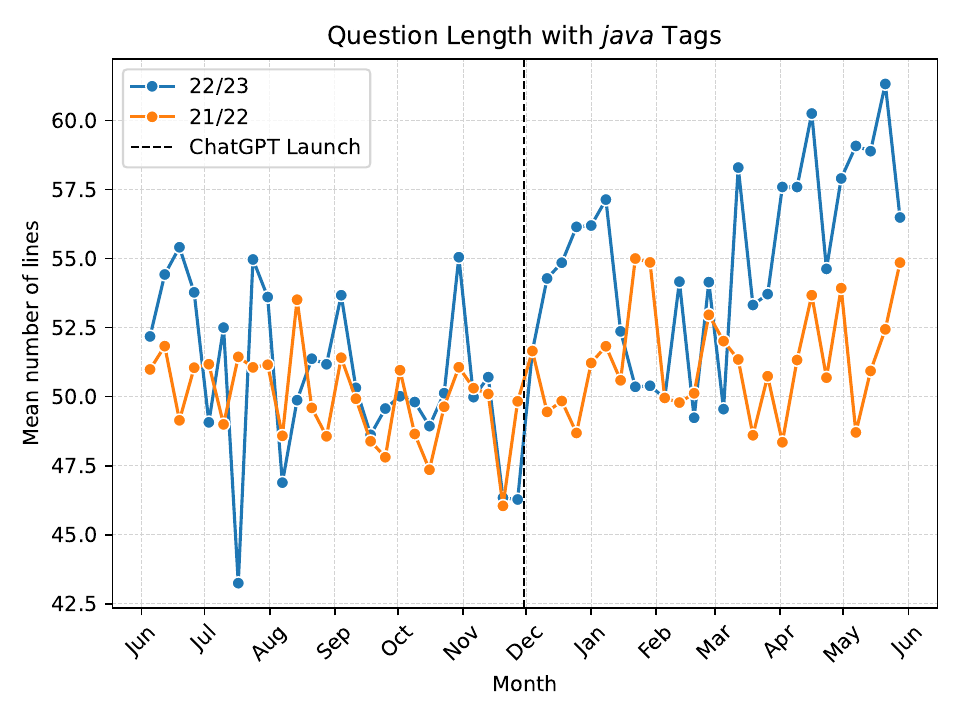}\label{fig:t_java_lc}}
    \subfloat[Java: lines of code]{\includegraphics[width=0.33\textwidth]{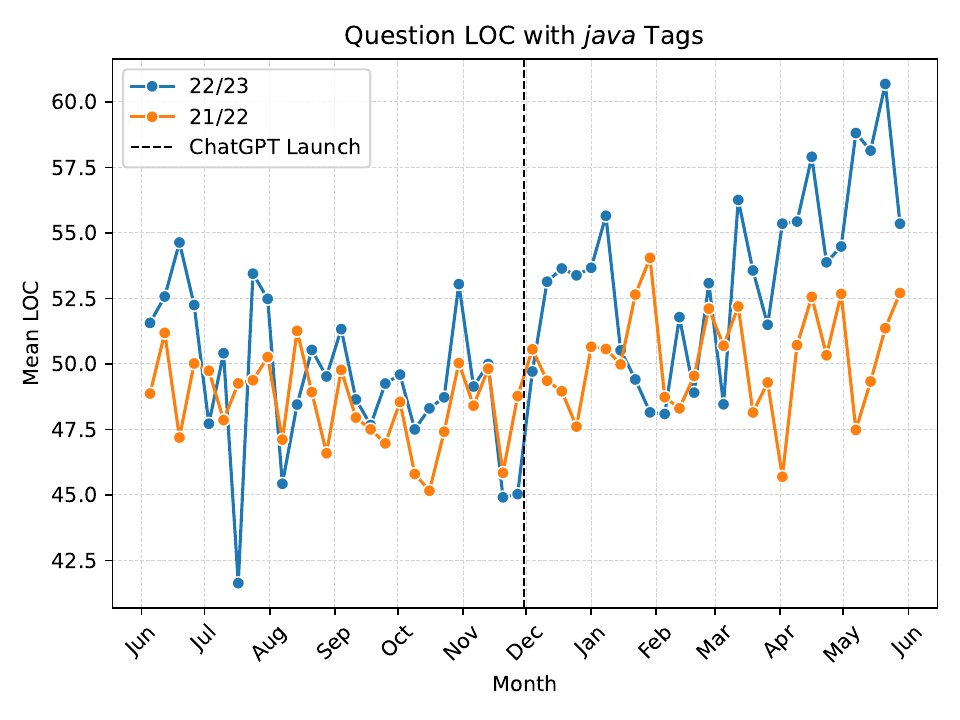}\label{fig:t_java_loc}}
	\caption{\textbf{Question tags on Stack Overflow.} We categorize questions by their tags (\textit{python} and \textit{java}) and tag groups (\textit{web} group includes \textit{javascript}, \textit{reactjs}, \textit{html}, \textit{node.js}, and \textit{css} tags) and show top three categories by question volume. In particular, we show \textit{web} tags in the first, \textit{python} tag in the second, and \textit{java} tag in the third row. Similar to the aggregate of all questions, we observe an accelerating downwards trend in the question volume in (a), (d), and (g), as well as a substantial positive trend in the length of questions in (b), (e), and (h), confirming the ChatGPT association across the disaggregated categories. In addition, we also observe a similar change towards positive trend in the length of the code examples in (c), (f), and (i) after the ChatGPT launch. We compute the lines of code by parsing the HTML body of the questions and extracting content from \textit{$<$code$>$} elements (used to format code examples on Stack Overflow). The positive trend in question and code length is somewhat weaker for smaller tags (\textit{java}) than for the larger tag groups (\textit{web} and \textit{python}). Nevertheless, we observe a similar positive trend in the remaining tag groups, which we do not show here due to limited space. This finding demonstrates that after the ChatGPT launch Stack Overflow users include longer code examples in their questions, suggesting a considerable shift in the problems for which users seek help on Stack Overflow.}
	\label{fig:raw}
\end{figure}

In Figure \ref{fig:tags} we show the top 10 tags in our \rev{two-year observation period (Fig. \ref{fig:ttc}), as well as top 10 tags from 22/23 (Fig. \ref{fig:tt}) and 21/22 (Fig. \ref{fig:tc}) only.}
The most frequent tag is \textit{python}, followed by two javascript related tags (\textit{javascript} and \textit{reactjs}), and \textit{java} tag. In all cases, we observe a skewed tag distribution as tag volume quickly falls off with higher ranks. The top 10 tags are quite stable over two periods with two tags dropping from top 10 in 22/23 (\textit{node.js} and \textit{pandas} are replaced by \textit{\rev{flutter}} and \textit{css}). 
To further analyze the question categories, we group \textit{javascript}, \textit{reactjs}, \textit{html}, \textit{node.js}, and \textit{css} tags into \textit{web} tag group as they thematically belong to the Web development. In Figures \ref{fig:t_web}, \ref{fig:t_python}, \ref{fig:t_java} we show the question volume of the top three tag groups and tags (\textit{web}, \textit{python}, and \textit{java}) and in all three cases, we observe a similar accelerating downwards trend as for aggregated questions and answers. The trend change is more prominent for larger question groups, i.e., \textit{web} and \textit{python} tags (see Figures \ref{fig:t_web} and \ref{fig:t_python}) have a stronger negative acceleration, while the trend change is weaker for \textit{java} (cf. Figure \ref{fig:t_java}). In addition to these three largest tags groups, we observe a similar relation between the intensity of trend change and the question volume for all other tags with smaller groups experiencing a weaker trend change, while larger question volumes generally tend to experience a higher acceleration of an already existing negative trend. Due to the space limitations we do not show here the figures for the remaining tag groups. In summary, we observe the same association between the acceleration of the negative trend in question and answer volume even after disaggregation of the data according to their thematic categories.

We continue and show the question length, measured in the number of lines, in Figures \ref{fig:t_web_lc}, \ref{fig:t_python_lc}, \ref{fig:t_java_lc}. Similar to the aggregated questions and answers, we observe the same association between the ChatGPT start and the question length in all tag groups. After the launch, we observe a substantial upwards tilt in the weekly averages of question length that persists until the end of our observation period. The tilt is more prominent for \textit{web} and \textit{python} tags (Figures \ref{fig:t_web_lc} and \ref{fig:t_python_lc}), but it is still clearly visible also for \textit{java} tag (Figure \ref{fig:t_java_lc}). Similar dependence of the intensity of the tilt on the question volume in a given group is visible in the remaining tags, not shown here due to space constraints. Hence, a clearly visible shift related to an increasing question length post-ChatGPT is also present in all tag groups.

\subsection{\rev{Lines of code}}
We also analyze the code length of examples included in the Stack Overflow questions. In the dataset, the post body is stored as HTML code, and the code examples are included verbatim within \textit{$<$code$>$} HTML elements. Hence, using an HTML parser, we extract all \textit{$<$code$>$} elements from the question body and count the lines included in those code snippets. We extract the code length separately for each tag group and plot its temporal development for top three tags in Figures \ref{fig:t_web_loc}, \ref{fig:t_python_loc}, and \ref{fig:t_java_loc}. In all three cases, we observe a similar development as with the overall question length. For larger tags, i.e., \textit{web} and \textit{python}, we see an extensive positive trend in code length after ChatGPT was introduced. In addition, we observe a similar, although somewhat weaker trend for a smaller \textit{java} tag, as well as for the remaining tags. This considerable discontinuity post-ChatGPT suggests a major change in help-seeking behavior for programming tasks on Stack Overflow. While the total number of questions across various thematic question groups further decreases, the length of the questions, as well as the length of code examples consistently and sustainably increases on Stack Overflow. Potentially, this indicates that users increasingly often turn to other channels such as ChatGPT for shorter, and to Stack Overflow for longer programming questions.

\begin{figure}[t]
	\centering
	\subfloat[Web: medium difficulty]{\includegraphics[width=0.33\textwidth]{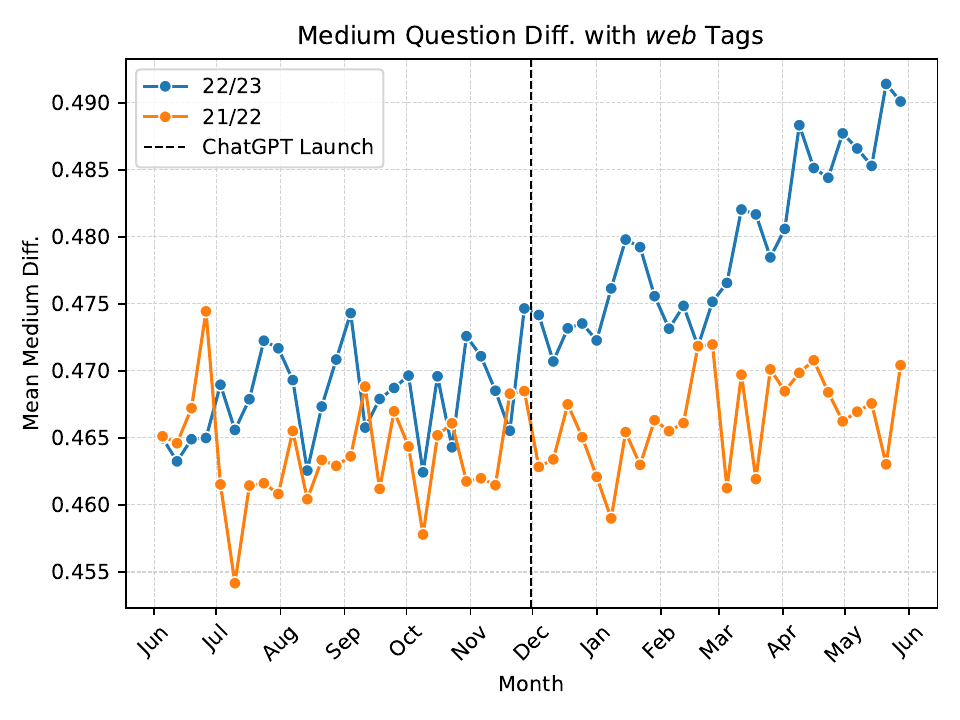}\label{fig:t_web_medium}}
	\subfloat[Python: medium difficulty]{\includegraphics[width=0.33\textwidth]{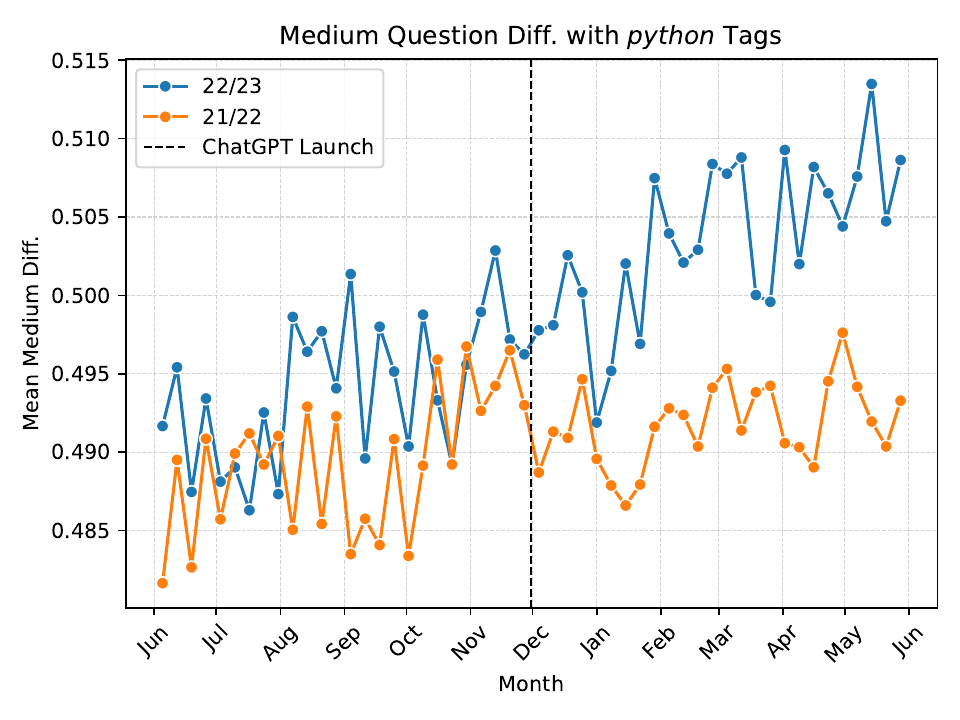}\label{fig:t_python_medium}}
    \subfloat[Java: medium difficulty]{\includegraphics[width=0.33\textwidth]{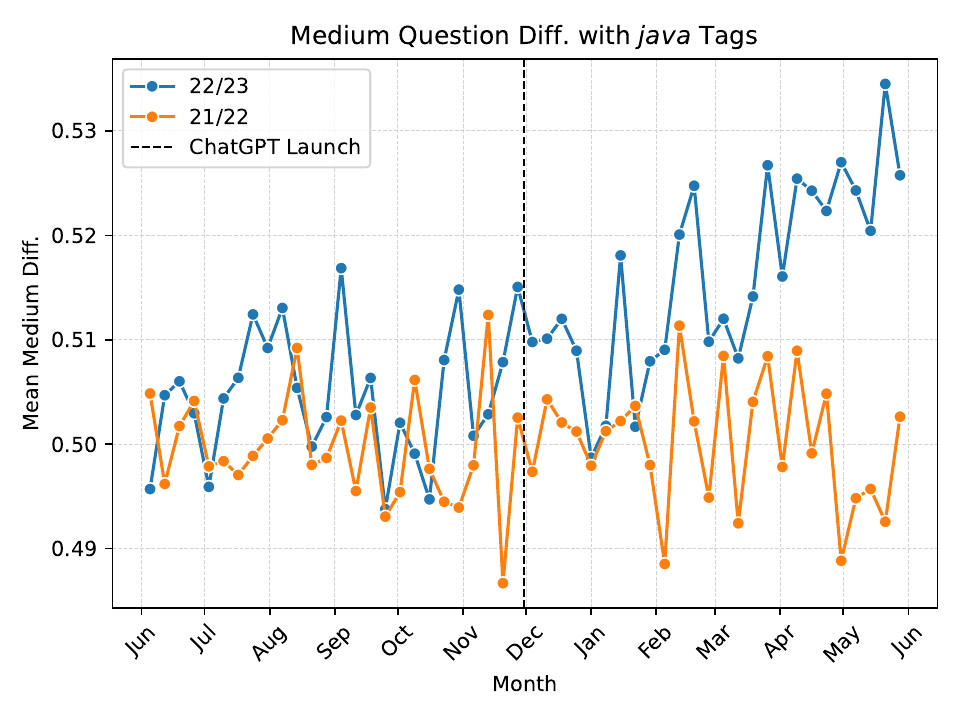}\label{fig:t_java_medium}}\\
    \subfloat[Web: hard difficulty]{\includegraphics[width=0.33\textwidth]{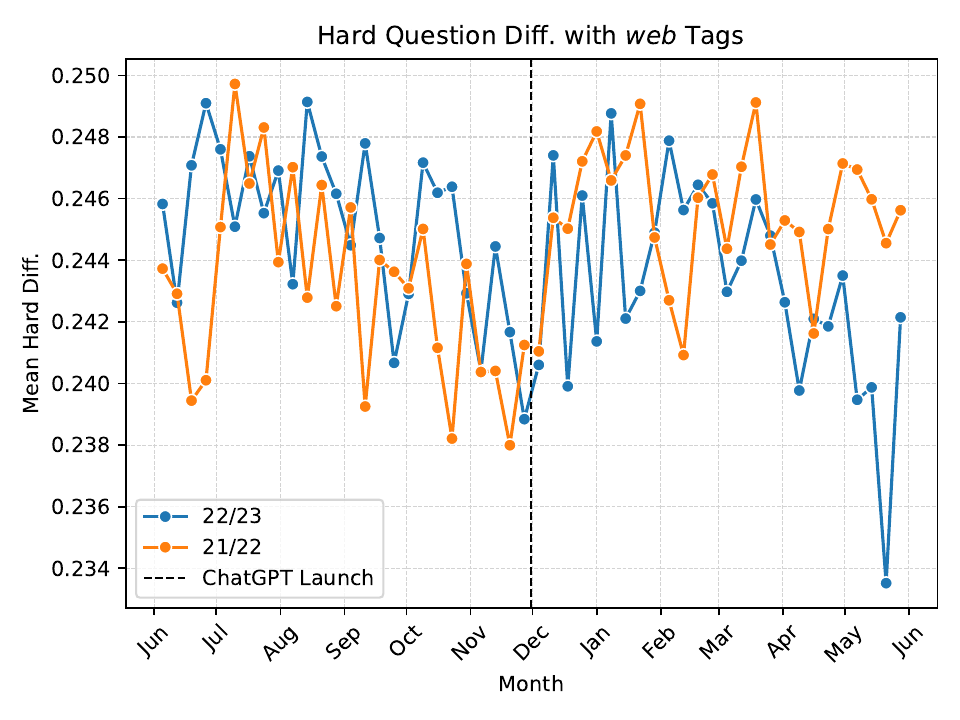}\label{fig:t_web_hard}}
	\subfloat[Python: hard difficulty]{\includegraphics[width=0.33\textwidth]{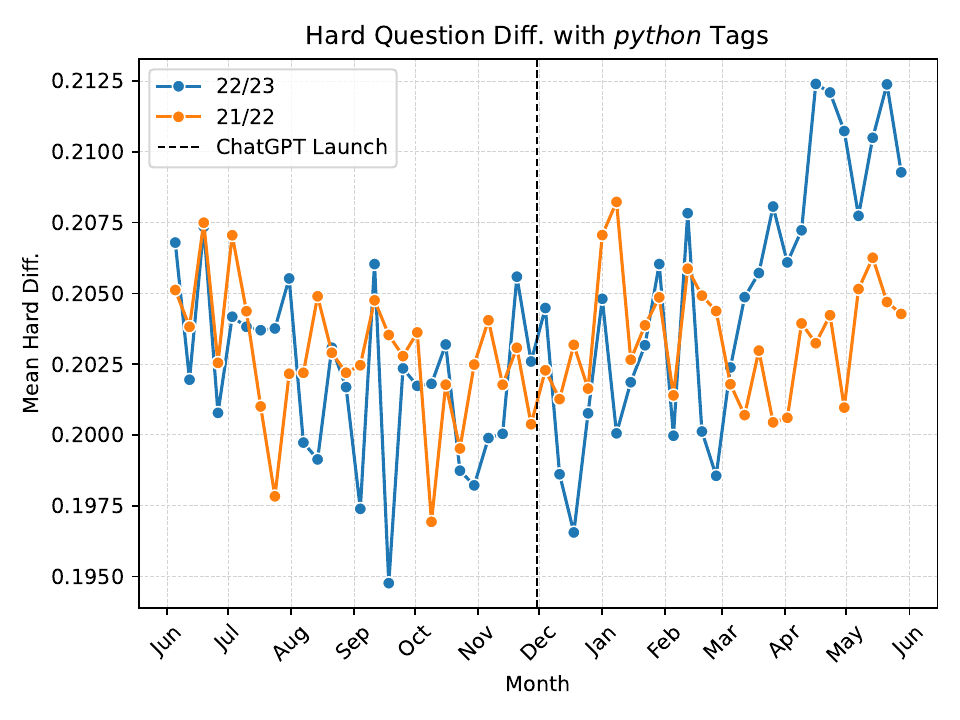}\label{fig:t_python_hard}}
    \subfloat[Java: hard difficulty]{\includegraphics[width=0.33\textwidth]{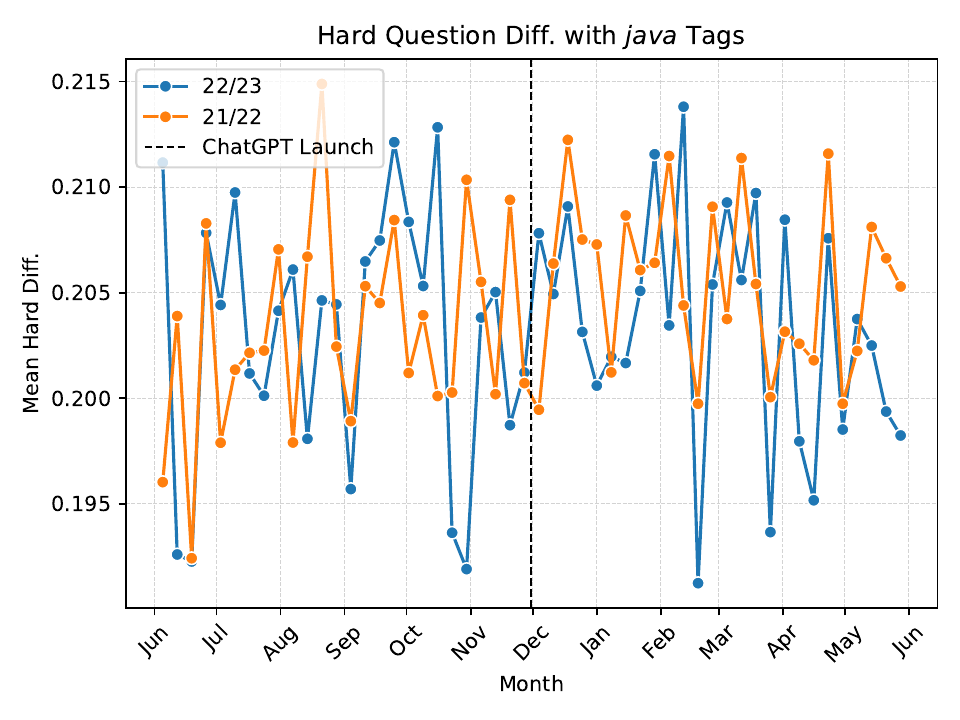}\label{fig:t_java_hard}}
	\caption{\textbf{Question and code \rev{difficulty} on Stack Overflow.} We classify questions by their difficulty as easy, medium, and hard. We first compute embeddings from questions including the code examples by using a pretrained question/code embedding model CodeT5 \citep{wang2021}. Using a labeled question/code dataset from LeetCode with easy, medium, and hard classes, we train an XGBoost classifier that we apply on the questions from Stack Overflow. The classifier computes the probabilities of questions belonging to given difficulty classes. In the top row, we show weekly means of the classifier probabilities for the medium difficulty on the top three question tag groups including \textit{web} (a), \textit{python} (b), and \textit{java} (c) tags. Similar to the question and code length for those tag groups we see an upward trend in the medium difficultly that accelerates around the ChatGPT launch suggesting a shift in user posting behavior in this period. The same analysis for the remaining tags, indicates, as previously with the question and code length, that smaller tag groups exhibit weaker changes in trends. In the bottom row, we depict the classifier probabilities for the hard difficulty. We do not observe any patterns or trend changes in these plots, as the probability for hard questions seems to develop completely at random, in particular for \textit{web} and \textit{java} tags in (d) and (f). In case of the \textit{python} tag in (e), we observe an upward trend starting in March 2023, already four months after the ChatGPT launch, hence, this change might be associated by some event other than introduction of ChatGPT.}
	\label{fig:difficulty}
\end{figure}

\subsection{\rev{Question and code difficulty}}
Finally, we analyze the difficulty of the Stack Overflow user questions together with the included code examples. To measure the question difficulty we first collect a dataset from LeetCode,\footnote{\url{https://leetcode.com/}, last retrieved on September 06, 2025.} which is an online platform collecting tasks for programmers' and software developers' training. The platform contains thousands of questions and programming tasks collected from programming courses and technical interviews in the industry. The questions are categorized into topics such as data structures, algorithms, or databases and by difficulty as being easy, medium, or hard tasks. Moreover, questions typically have solutions in several programming languages including \textit{python}, \textit{c++}, or \textit{javascript}. We collect a prepared LeetCode dataset\footnote{\url{https://huggingface.co/datasets/greengerong/leetcode}, last retrieved on September 06, 2025.} from huggingface.\footnote{\url{https://huggingface.co/}, last retrieved on September 06, 2025.} The dataset includes $2,360$ programming tasks with examples for all three difficulty levels together with solutions in \textit{python}, \textit{java}, \textit{c++}, and \textit{javascript}. For each task and a solution language, we create a separate instance containing the title of the question, the description of the task, and the solution in that language, resulting in the final dataset with $9,440$ examples. Second, we use a pretrained text/code encoder transformer model CodeT5 \citep{wang2021} that was trained on the CodeSearchNet dataset \citep{husain2020}, a large collection of text (query-like questions) and functions in several programming languages.\footnote{\url{https://github.com/github/CodeSearchNet}, last retrieved on September 06, 2025.} With this model, we embed the examples from our LeetCode dataset. In particular, we use CodeT5-Base\footnote{\url{https://huggingface.co/Salesforce/codet5-base}, last retrieved on September 06, 2025.} from huggingface that computes 768-dimensional embeddings of the input text/code content. \rev{We prepare the input for the embedding procedure by concatenating the task title prefixed by \textit{``Problem:''}, the task description, and the solution code prefixed by \textit{``Code(language):''}, where language corresponds to the current programming language.}
Third, using the computed embeddings as the input features we train an XGBoost classifier \citep{chen2016} with the question difficulty as the classification target. We split the LeetCode dataset in 80\% training and 20\% test dataset and train the classifier using the default parameters.\footnote{\url{https://xgboost.readthedocs.io/en/stable/python/python_intro.html}, last retrieved on September 06, 2025.} On the test dataset, we achieve ROC-AUC of $0.99$ and weighted macro F1-average of $0.95$. Fourth, as we achieve high accuracy rates on the test dataset, we retrain the XGBoost classifier on the whole dataset using again the default classifier parameters. Fifth, we embed all questions from our Stack Overflow dataset by \rev{concatenating} the question title, the question body, the tag as an indication of the programming language, and the code examples \rev{in the same way as the examples from LeetCode}. Finally, we predict the question difficulty with our XGBoost classifier to obtain the probabilities that a given question belongs to the easy, medium, or hard difficulty class.

We depict the temporal evolution of the predicted question difficulty (as the weekly means of the probability of the medium and hard difficulty) for the three largest tag groups and for both of our observation periods in Figure \ref{fig:difficulty}. In the top row (Figures \ref{fig:t_web_medium}, \ref{fig:t_python_medium}, and \ref{fig:t_java_medium}), we show the weekly mean probability for the medium difficulty and in the bottom row (Figures \ref{fig:t_web_hard}, \ref{fig:t_python_hard}, and \ref{fig:t_java_hard}) the weekly mean probability for the hard difficulty. For the medium difficulty, we observe a significant upward trend around and after the ChatGPT start in all three tag groups. This indicates a substantial shift in the user posting behavior post-ChatGPT associated with the difficulty of the questions. This increase in probability of the medium difficulty comes at the cost of the probability of easy questions (which we do not show here) as the probabilities of the hard questions are quite noisy but without any visible trends (cf. bottom row in Figure \ref{fig:difficulty}). Hence, on Stack Overflow in the post-ChatGPT period, the probability of medium difficulty questions increases while the probability of easy questions decreases without any visible changes in the probability of hard questions. Potentially, this indicates that the users turn more frequently to other channels such as ChatGPT for easy programming questions, but ask community more often in the case of more difficult questions. However, users seem to further differentiate in their reliance on the community response---in cases of the medium problem difficulty users turn more often to the community, while in cases of the hardest questions no significant difference in behavior is visible.

\section{Effect of ChatGPT on Stack Overflow}

\subsection{\rev{Difference-in-differences model}}
To quantify the effect of ChatGPT on Stack Overflow we use a difference-in-differences (DiD) regression setup. DiD analysis allows us to estimate the ChatGPT effect on various Stack Overflow metrics while controlling for temporal trends. In particular, we take the 21/22 period as a control group and compare temporal developments of the period 22/23 before and after the ChatGPT launch to that control group. We run our DiD regression for multiple outcome variables including the question and answer length, question and answer scores and views, the length of questions and code examples for all tag groups, as well as question difficulty estimates.

Hence, denoting our outcome variable with $Y_i$, we model this variable as a linear function of period ($P$) denoting whether the timestamp of the question is before the ChatGPT launch or after (we use one year prior to the launch, i.e., November 30, 2021 as the event date for the control period), ChatGPT treatment ($T$) denoting whether the period is 22/23 (treated) or 21/22 (control) and the interaction between $P$ and $T$. The interaction coefficient is the DiD estimate and it quantifies the change in slope of the linear function of the outcome variable as an effect of the ChatGPT launch. \rev{In addition, to account for seasonal (see next subsection, Sec. \ref{sec:seasonality}) and observed temporal trends (cf. Figures \ref{fig:questions}, \ref{fig:answers}, \ref{fig:raw}, and \ref{fig:difficulty}), we use the week of the year ($W$), and the day of the week ($D$) as control variables. We include both of these control variables as categorical variables, effectively capturing seasonality within the calender week ($W$) as well as any potential weekday patterns ($D$). This enables a comparison of outcome variables within the same weekday and same calendar week, in the treated and the control year.} 
Our final DiD model is given by:
\rev{
\begin{equation}
    Y_i = \beta_0 + \beta_1 T + \beta_2 P + \beta_3 (T \cdot P) + \beta_4 W + \beta_5 D + \epsilon_i,
    \label{eq:did}
\end{equation}
}
where $\beta_1$ quantifies the average change in prediction of the outcome variable between the 21/22 and 22/23 periods, $\beta_2$ the average change in the outcome variable before and after the launch (or before and after Novemeber 30, 2021 for the 21/22 period), and $\beta_3$ is our DiD estimate that quantifies the effect of ChatGPT on a given Stack Overflow metric after controlling for temporal trends. \rev{Finally, $\beta_4$ and $\beta_5$ quantify any residual weekly and daily trends in a given metric.}

\subsubsection{\rev{Seasonality effects}}\label{sec:seasonality}
\rev{Similar to other cases of user-generated data, we expect that the user posting behavior is correlated over time, for example, over subsequent days or weeks. Therefore, we substantiate the visual inspection of temporal trends in our data (cf. Figures \ref{fig:questions}, \ref{fig:answers}, \ref{fig:raw}, and \ref{fig:difficulty}) by quantifying the seasonal effects with the autocorrelation function \cite{bertrand2004}. In particular, for each outcome variable and the data selection, we (i) fit our DiD regression (Eq. \ref{eq:did}), (ii) compute the residuals, and (iii) inspect how these residuals are correlated over time for a maximum lag of $60$ days. Our analysis indeed reveals moderate week-level autocorrelations, suggesting weekly cycles of posting behavior as for the majority of the outcome variables the autocorrelation reaches the correlation threshold only on the first and seventh day. Hence, including the day of the week and the week of the year as categorical variables in our DiD model accounts for systematic seasonal weekday versus weekend patterns, as well as weakly seasonal differences over the whole year.}

\subsubsection{\rev{Parallel trend assumption}}
Difference-in-differences models assume a parallel trend between the control and treatment groups in the period before treatment. We check for this trend visually (cf. Figures \ref{fig:questions}, \ref{fig:answers}, \ref{fig:raw}, and \ref{fig:difficulty}, blue vs. orange lines in before ChatGPT periods) and see similar and stable trends in before treatment period. To provide further evidence for parallel trend assumption we additionally fit the following regression model:
\rev{
\begin{equation}
    Y_i = \beta_0 + \beta_1 T + \beta_2 (T \cdot t) + \beta_3 W + \beta_4 D + \epsilon_i,
    \label{eq:trends}
\end{equation}}\rev{
where as before,  $T$ denotes treated vs. control group, $W$ the week of the year, $D$ the day of the week (categorical variables), and $t$ is the week counter in a given year (an integer variable). We fit this model only with the data from the period before the treatment begins, i.e., data with $P=0$. Coefficient $\beta_2$ estimates the difference in slopes between the treated and the control group, and the test whether $\beta_2=0$ provides information on whether trends between two groups are different prior to treatment.}

\subsubsection{\rev{Distributions of the outcome variables}}
Before computing DiD regressions, we check the distributions of the outcome variables. Similar to other user-generated data, we find highly skewed distributions for several outcome variables such as question and answer length, as well as code examples length across all tag groups. In these cases, majority of questions and answers have small values of the outcome variable and a much smaller portion of questions and answers have large values for those particular quantities. Therefore, we log transform all of these outcome variables to obtain more symmetric distributions closer to a normal distribution. We leave other outcome variables without significant distributional skew such as scores or difficulty probability unchanged. The log transformation of the outcome variable turns these regression models into multiplicative models and changes the interpretation of the coefficients. In particular, coefficients close to zero, e.g., 0.02 can be interpreted directly as the 2\% change in the outcome variable for one unit change in the corresponding variable while all other variables are being held constant, with the exact percentage change being equal to $(e^\beta - 1)\times100\%$ for coefficient values not close to zero. We discuss the particular interpretation of the corresponding regression models directly in the results section.

\subsubsection{\rev{Regression setup}}
To properly estimate short-term and long-term effects of the ChatGPT launch, as well as the sensitivity of the exact launch date on Stack Overflow, we adopt the following DiD setup. We fit multiple regression models, always taking all the data from the pre-launch period into regression analysis, i.e., data where $P=0$, which includes six months of data from May 31, 2021 until November 29, 2021 for the control group and six months of data from May 30, 2022 until November 29, 2022 for the treated group. In addition, we use a sliding window of one month of data in the post-launch period and iterate over the post-launch data starting with November 30, 2021 respectively November 30, 2022 until \rev{May 29}, 2022 respectively \rev{May 28}, 2023. In total, we fit 151 regression models for each metric and obtain a daily time series of DiD coefficients starting with the ChatGPT launch and extending for six months in future, allowing us to estimate the short-term effects with data selections closer to the launch, as well as, long-term effects with data selections further from the launch. This ``secret weapon'' methodology, introduced by Gelman and Huang \citep{Gelman2008}, allows us to quantify effect trends rather than mere point estimates. 

As a further preprocessing step of our regression setup, we standardize all the outcome variables such that the regression coefficients are measured on the scale of the standard deviation of a given metric. This makes the regression results comparable between different outcome variables. 
\rev{Futhermore, as our analysis of the seasonal effects suggests a moderate weekly cycle, in all our regressions we estimate the standard errors with a heteroskedasticity- and autocorrelation-consistent (HAC) robust estimator. In particular, we use Newey-West covariance matrix estimator \cite{newey1987} with the maximal lag of seven days to account for the empirical autocorrelation function. Finally, we conduct a series of additional stability and sensitivity checks to estimate the robustness of our main regression results. We describe these robustness experiments and their results in Section \ref{sec:robust}.}

\subsection{Regression results}

\subsubsection{\rev{Parallel trends}}
Following the Equation \ref{eq:trends}, we fit regression models for all combinations of the outcome variables and our data groups including the aggregated data and the data separated by the tags. The \rev{$\beta_2$} coefficient indicating the difference in trends between the control and the treated group prior to ChatGPT introduction is close to zero in all cases. Specifically, for the aggregated data the largest magnitude of \rev{$\beta_2$ is $-0.0024$ (p-value $<0.0001$) for question views with all other estimated coefficients being closer to zero. For the three largest tag groups, we obtain $-0.002$ (p-value $< 0.0001$) for \textit{web} question views, $-0.0029$ (p-value $<0.0001$) for \textit{python} question scores, and $-0.0027$ (p-value $<0.0001$) for \textit{java} question views as the largest magnitudes of $\beta_2$ coefficient}, with coefficients in all other cases being even closer to zero. Hence, apart from the visual inspection of temporal plots of the quantities in question, these tests provide further evidence in favor of parallel trend assumption. 

\begin{figure}[t]
	\centering
	\subfloat[Question length]{\includegraphics[width=\textwidth]{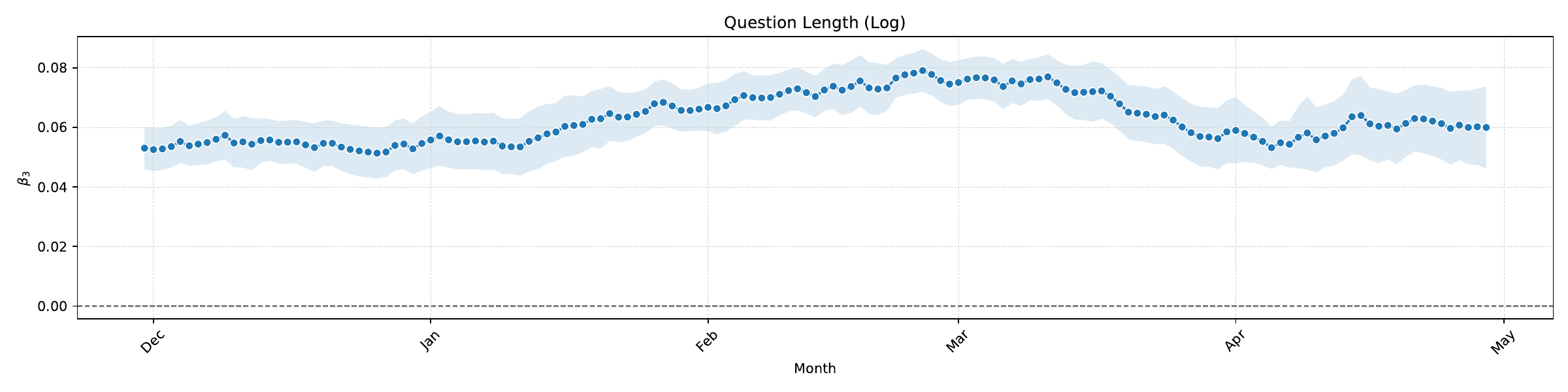}\label{fig:qlc}}\\
    \subfloat[Answer length]{\includegraphics[width=\textwidth]{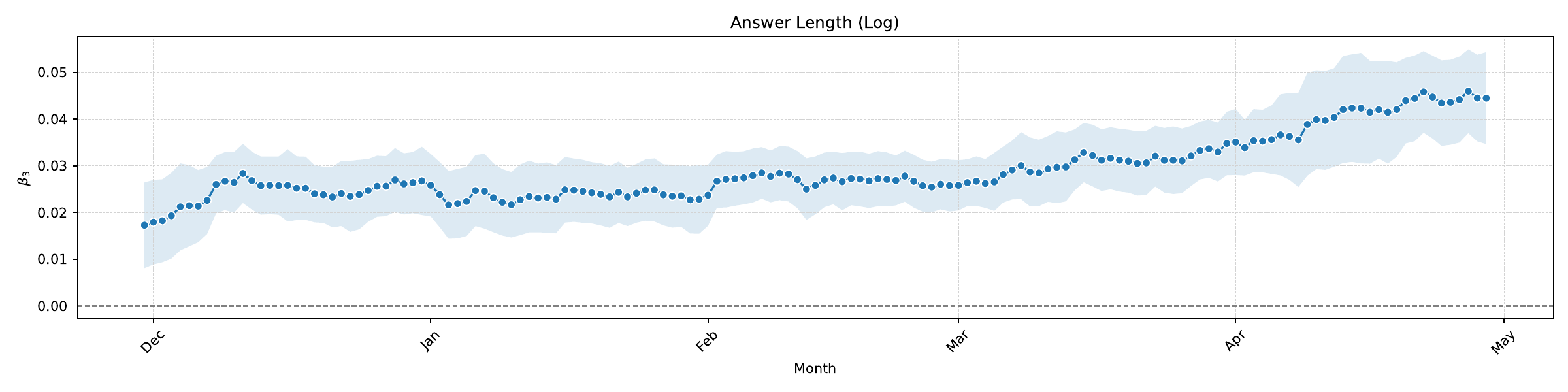}\label{fig:alc}}
	\caption{\textbf{Effect of ChatGPT on the question and answer length over time.} We fit multiple DiD regressions using periods 22/23 as the treated group and 21/22 as the control group. We fit our models using all the data prior to the ChatGPT launch (or prior to November 30, 2021 for the control group) and one month of data after the launch, starting with the launch date and iterating with a daily sliding window until the end of our observation period. In total, for every outcome variable we fit 151 regression models and plot the $\beta_3$ coefficient as our DiD estimate, quantifying the effect of ChatGPT on the particular outcome variable. We accompany the estimates with the confidence intervals ($\pm2\times se$) and check for intersection of those confidence intervals with the zero line. We interpret the estimates without intersection as statistically significant effects of ChatGPT on the given quantity. For the question length, depicted in (a), we observe a strong effect that persists over complete six months observation period. Similarly, in (b) we observe a strong positive and continuous effect of ChatGPT on the answer length. We conclude that, in addition to ChatGPT negatively affecting the volume of question and answers on Stack Overflow \citep{burtch2024, delrio2024}, it had, at the same time, a profound effect on the user content length, resulting in longer user questions as well as answers from the community.}
	\label{fig:did_qa}
\end{figure}

\subsubsection{\rev{Interpretation of the results}}
\begin{figure}[t!]
	\centering
	\subfloat[Web: question length]{\includegraphics[width=\textwidth]{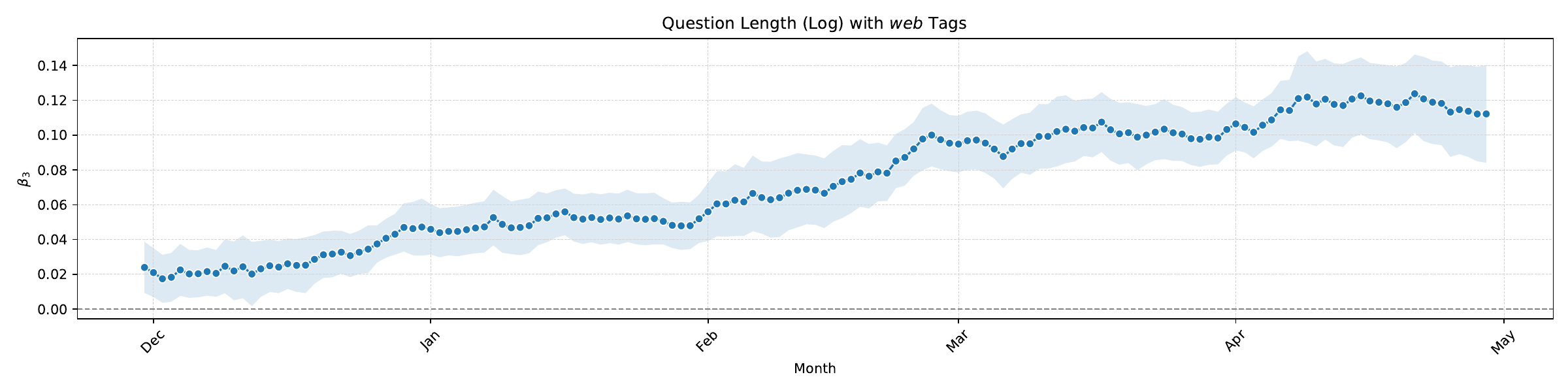}\label{fig:web_qlc}}\\
    \subfloat[Python: question length]{\includegraphics[width=\textwidth]{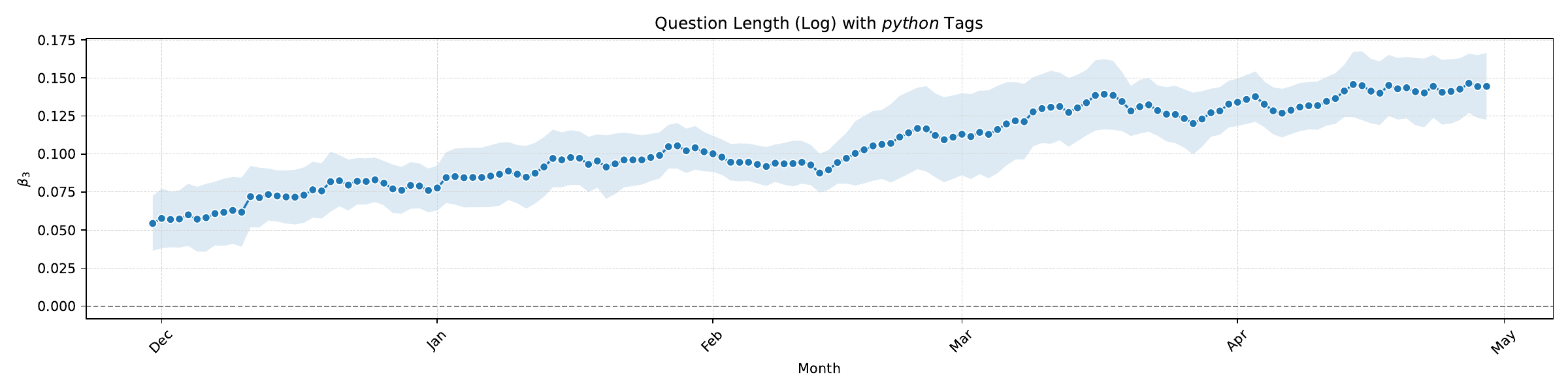}\label{fig:python_qlc}}\\
    \subfloat[Java: question length]{\includegraphics[width=\textwidth]{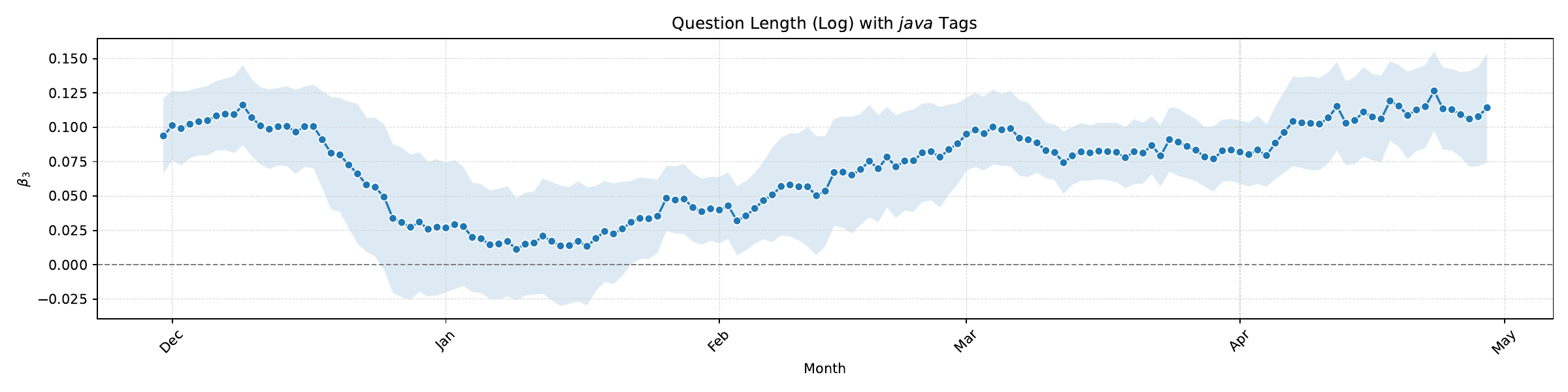}\label{fig:java_qlc}}
	\caption{\textbf{Effect of ChatGPT on the question length over tags and time.} Using the same regression setup as before, we depict the effect of ChatGPT on question length across tag groups. We observe strong positive and lasting effects on the question length for \textit{web} tags in (a), \textit{python} tag in (b) and \textit{java} tag in (c). In particular, the effect sizes at the end of the observation period in May 2023 are $12\%$ for \textit{web} tag, $16\%$ for \textit{python} tag, and $9\%$ for \textit{java} tag of the individual standard deviations of the question length. We also observe a slightly smaller trend for the smallest \textit{java} tag than for large volume \textit{web} and \textit{python} tags. By dividing the data into thematic categories we provide further evidence for a foundational change in the Stack Overflow's Q\&A practices since the introduction of ChatGPT by ruling out the possibility of Simpson's paradox \citep{blyth1972}, in which an association or effect observed in the aggreagated data disappears once when data is dividied into categories.}
	\label{fig:did_tags}
\end{figure}

\begin{figure}[t!]
	\centering
    \subfloat[Web: code length]{\includegraphics[width=\textwidth]{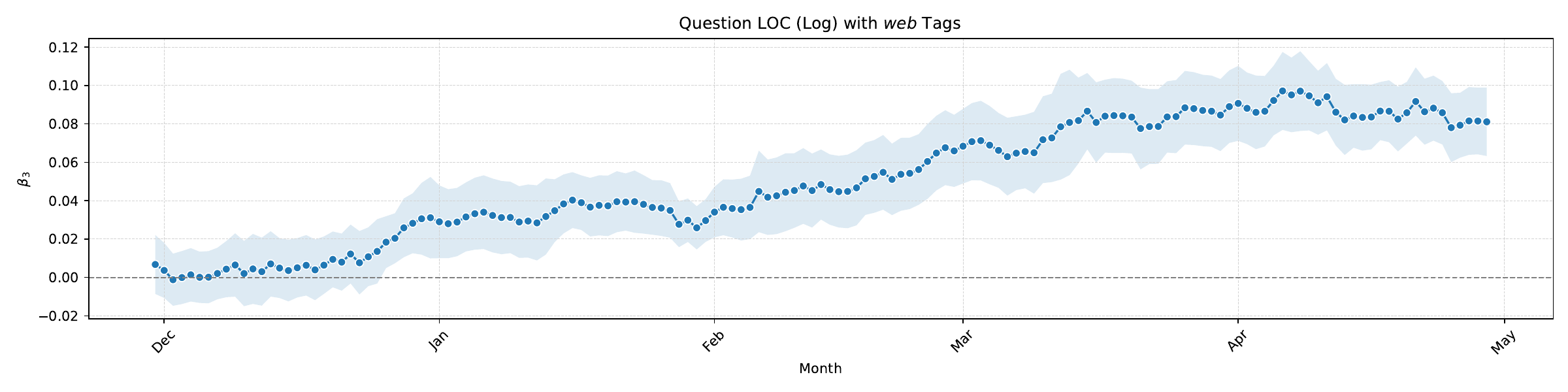}\label{fig:web_loc}}\\
    \subfloat[Python: code length]{\includegraphics[width=\textwidth]{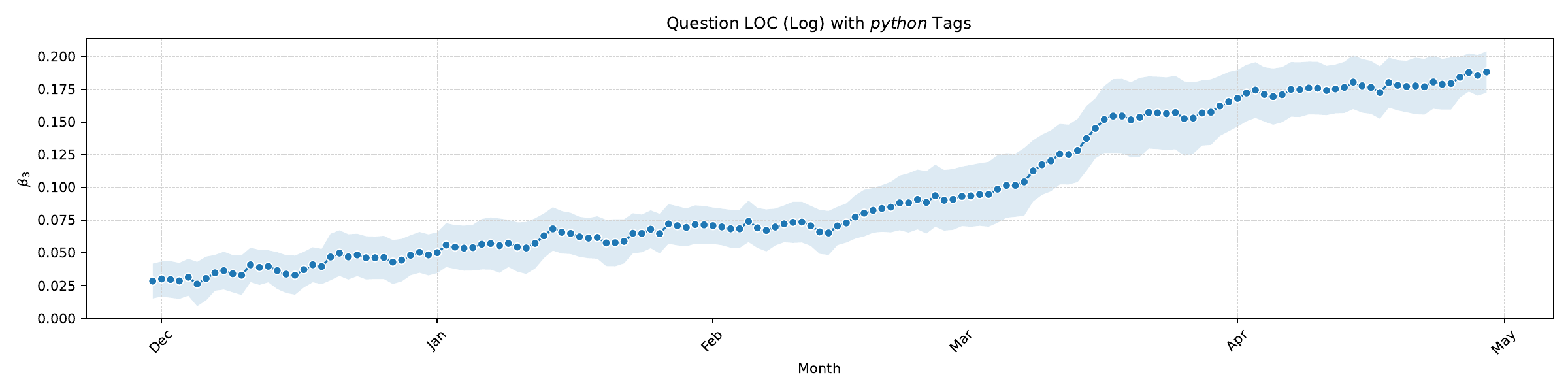}\label{fig:python_loc}}\\
    \subfloat[Java: code length]{\includegraphics[width=\textwidth]{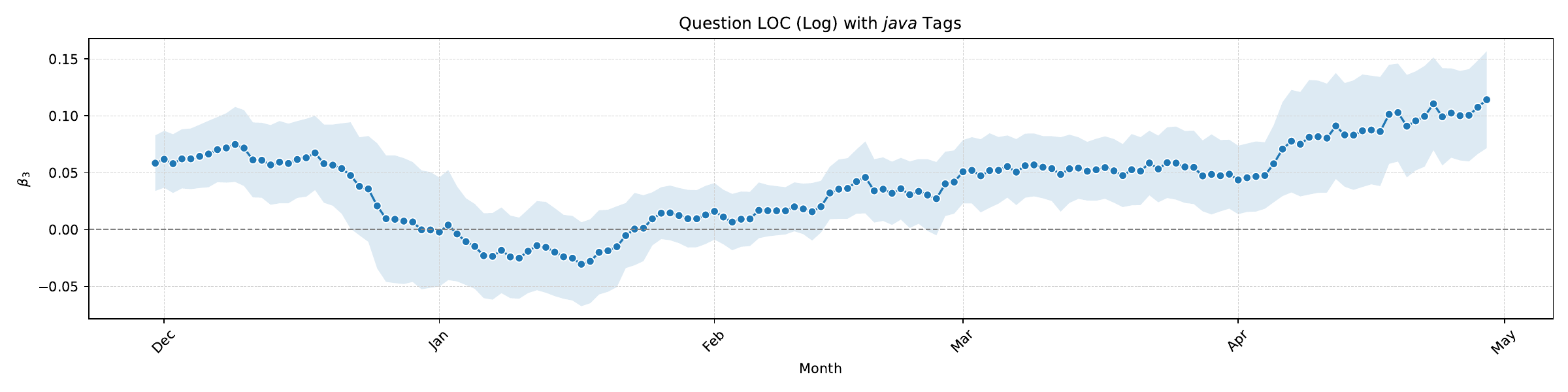}\label{fig:java_loc}}
	\caption{\textbf{Effect of ChatGPT on the code length over programming language and time.} Again, utilizing our DiD setup we estimate the effect of ChatGPT on the code length users post in their code examples on Stack Overflow. Similar to the question length, we observe a strong and consistent positive effect for \textit{web} tag in (a) and \textit{python} tag in (b). In case of \textit{java} tag in (c), the effect oscillates in the first few months until it reaches similar sizes as for the other two tags in the last month of our observation period. While larger tag groups experience larger effect sizes, for all three programming languages the example code length increases substantially in post-ChatGPT period.}
	\label{fig:did_tags_loc}
\end{figure}

In Figures \ref{fig:did_qa}, \ref{fig:did_tags}, \ref{fig:did_tags_loc}. and \ref{fig:did_tags_medium}, we show the \rev{main} results of our DiD regressions for question and answer lengths, question and answer lengths for three largest tag groups, code length from those largest tag groups, and question difficulty, respectively. In all figures, we show the temporal development of the ChatGPT effect as measured by the interaction coefficient $\beta_3$ from Equation \ref{eq:did}, together with the confidence intervals (blue shaded region around the DiD estimate) computed as $\pm2\times se$, $se$ being the standard error of the DiD coefficient estimate. In cases where confidence intervals do not intersect the zero line, the coefficient is statically significant, indicating that the ChatGPT launch had an effect on the given quantity. The coefficent values can be interpreted as the percentage change (in Figures \ref{fig:did_qa}, \ref{fig:did_tags}, \ref{fig:did_tags_loc}) or as the increase in probability of the outcome variable (in Figure \ref{fig:did_tags_medium}) measured at the scale of the standard deviation of that variable. This setup allows for comparison of the effect size between different outcome variables.

\subsubsection{\rev{Question and answer length}}
In Figure \ref{fig:did_qa}, we observe a positive, significant DiD coefficient over the whole observation period, indicating a persistent positive effect of ChatGPT on the length of the questions and corresponding answers after ChatGPT was introduced. The effect is stronger for questions and oscillates between $6\%$ and $8\%$ of the standard deviation of the question length, and amounts to an increase of $2\%$ to $3\%$ of the standard deviation of the answer length, with a slight upwards trend towards the end of the observation period, reaching almost $5\%$ increase in the end signaling a persisting long term effect of ChatGPT on question and answer length on Stack Overflow.

\subsubsection{\rev{Tags}}
In Figure \ref{fig:did_tags}, we show the DiD results for question length after dividing the data according to their tag groups. We observe almost identical results as before. In all tag groups, DiD coefficients are positive and significant during whole observation period, except for a few days in the beginning of December 2022 for \textit{web} tags, as well as approximately two months period between end of December 2022 and beginning of February 2023 for \textit{java} tag, where the confidence intervals intersect with the zero line although the estimated effect remains positive at all times. Apart from that, we observe a strong positive effect with an upwards trend for \textit{web} and, in particular, \textit{python} tag, and a slightly oscillating estimate in the beginning of the observation period for \textit{java} tag, which also stabilizes at high positive values around February 2023. The effect sizes are higher for three largest tag groups than for the aggregated data and we corroborate this with the DiD results for the remaining tag groups where we observe stronger effect sizes for larger tags and smaller for tags with a smaller number of questions (due to the space restrictions we do not show these results here). In particular, the effect sizes at the end of the observation period are around $12\%$ for \textit{web} tag, $16\%$ for \textit{python} tag, and $9\%$ for \textit{java} tag (measured at the scale of the individual standard deviations of the question length), again signaling a strong and consistent long term effect of ChatGPT on the question length across the tags.

\begin{figure}[t!]
	\centering
    \subfloat[Web: medium difficulty]{\includegraphics[width=\textwidth]{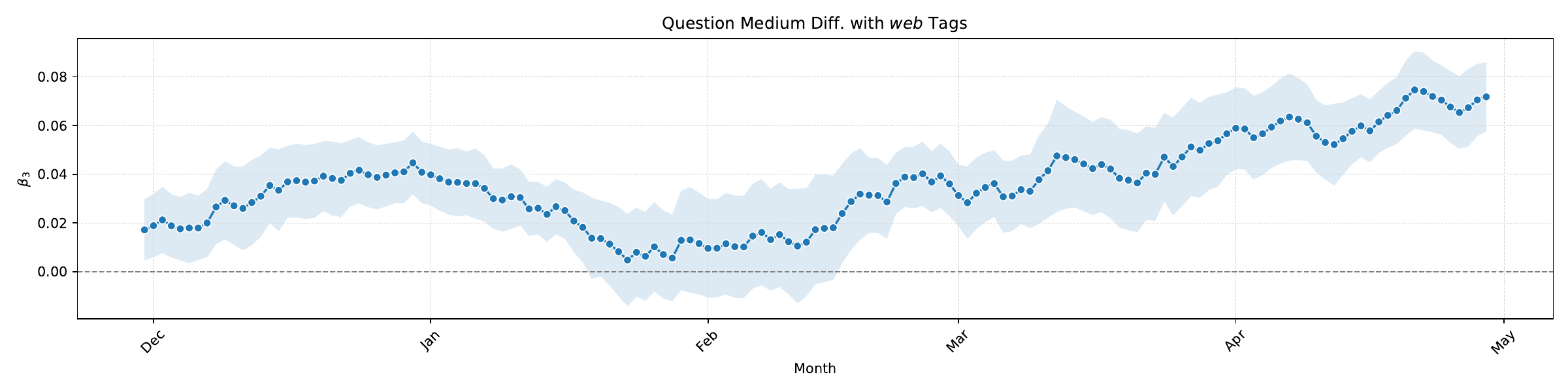}\label{fig:web_medium}}\\
    \subfloat[Python: medium difficulty]{\includegraphics[width=\textwidth]{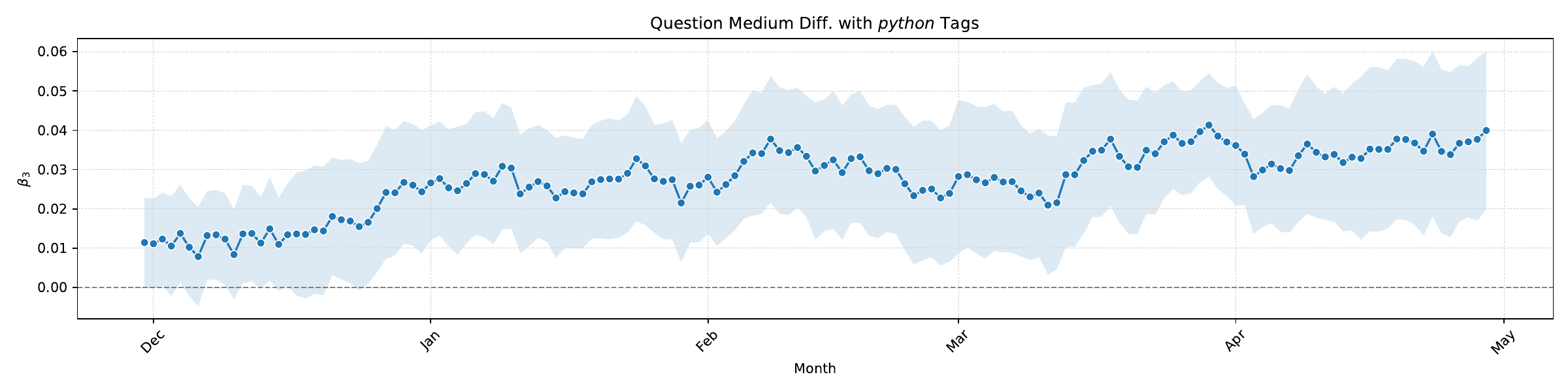}\label{fig:python_medium}}\\
    \subfloat[Java: medium difficulty]{\includegraphics[width=\textwidth]{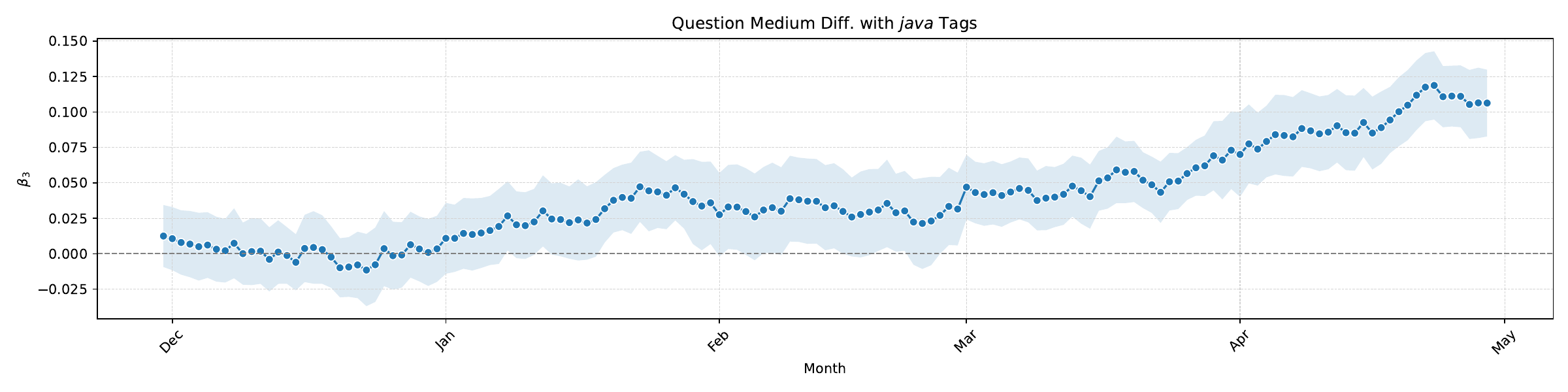}\label{fig:java_medium}}
	\caption{\textbf{Effect of ChatGPT on the question difficulty over programming language and time.} With our DiD setup, we estimate the effect of ChatGPT on the question difficulty (here the probability of the medium question difficulty) on Stack Overflow. Similar to the question length and code length, we observe a substantial upward trend in the DiD estimate for \textit{web} tag in (a) and \textit{python} tag in (b) in the first two months after the introduction of ChatGPT. While the trend remain stable for \textit{python} tag, there is a short period of time  (around February 2023), in which the DiD estimate is not significant for \textit{web} tag, before it again tilts upwards reaching almost 8\% of increase in the probability of medium question difficulty at the end of the observation period. In case of \textit{java} tag in (c), the effect oscillates around zero in the first month until it reaches sizes similar to \textit{python} and \textit{web} tag in May 2023. As previously, we observe that larger tag groups experience larger effect sizes.}
	\label{fig:did_tags_medium}
\end{figure}

\subsubsection{\rev{Code length}}
In Figure \ref{fig:did_tags_loc}, we depict the DiD regression results for the code length of examples included in the questions that we divide according to their tags. These results almost completely mirror the results of the question length for specific tags. In particular, for \textit{python} tag we observe a consistent positive effect with an increasing trend. The effect size at the end of the observation period is around $21\%$ increase in the python code examples length. For the \textit{web} tag after a short period of approximately one month after the start of ChatGPT, the coefficient increases and remains positive and significant until the end of the observation period. The final effect sizes is around $8\%$. In case of the \textit{java} examples, the coefficient oscillates in the first few months, but reaches a steady positive value from March 2023 onward. The final effect size is around $12\%$ of the standard deviation of the \textit{java} code examples. Similar to the question length for the individual tags, we compute DiD coefficients for all remaining tags and observe that, in general, the effect sizes are higher for larger tags (we do not show these results here due to space limitations). In summary, our results suggest a persistent and a strong positive effect of ChatGPT on the length of the code examples included in the questions on Stack Overflow.

\subsubsection{\rev{Question and code difficulty}}
Finally, in Figure \ref{fig:did_tags_medium} we show the temporal evolution of the DiD estimate for the XGBoost classifier probability of the question medium difficulty for three largest tag groups. Again, we standardize the outcome variables such that the results can be interpreted as the increase in the probability of predicting medium question difficulty measured at the scale of the standard deviation of that probability. We observe, a consistent and positive trend for the DiD estimate over the whole observation period. There are slight oscillations for the \textit{web} tag (see Figure \ref{fig:web_medium}) around February 2023, and for the \textit{java} tag in the first month of the observation period, but overall the DiD estimate is consistently significant and positive in our observation period. Again, the effect sizes is associated with the post volume, as our calculation with the smaller tags show smaller effect sizes or inconsistent and non-significant DiD coefficients for prolonged periods of time (we omit these plots from the paper due to space limitations). However, the effect sizes in May 2023 for the three largest tag groups are around 7\% for \textit{web}, 4\% for \textit{python}, and 11\% for \textit{java} tag of the standard deviation of the medium difficulty probability. This suggests a strong shift towards more difficult questions posted on Stack Overflow when compared to the pre ChatGPT period. The coefficients for the hard question difficulty are noisy and non-significant for longer periods suggesting that Stack Overflow users increasingly posted questions of the medium difficulty at the expenses of easy questions after the ChatGPT launch.

\subsubsection{\rev{Remaining outcome variables}}
We note here that we also fitted DiD regressions for further outcome variables such as question views, question scores, and answer scores for the aggregated data as well as for tag groups. We do not find any consistent effects for most of these outcome variables except for a strong negative effect on question views for the aggregated data and the largest tag groups such as \textit{web} and \textit{python}, confirming once more results from the previous studies \citep{burtch2024, delrio2024} on the declining question volume post-ChatGPT.

\subsection{\rev{Robustness analysis}}\label{sec:robust}

\rev{To estimate the robustness of our results, we perform a series of sensitivity and stability computations. Apart from multiple regression fits with a rolling one-month window in the post-ChatGPT period (``secret weapon''), we introduce three results summary statistics that we compute from the regression results. These summary statistics include:
\begin{enumerate}[(i)]
 \item \textit{Longest positive effect streak (LPES).} Using $151$ regression coefficients for the DiD estimator ($\beta_3$, Eq. \ref{eq:did}) and their lower confidence bound ($-2\times se$), we extract the longest continuous interval where the lower confidence bound is greater than zero ($\beta_3 - 2 \times se > 0$), and take the length of this interval in days as LPES. Hence, LPES captures the persistence of the effect over our six months post-ChatGPT observation period.
 \item \textit{Maximum of the rolling mean (MRM).} Following our one month rolling window regression setup, we compute the mean effect size for each rolling window and then take the maximum of these $151$ rolling windows as MRM to capture the peak effect size in the observation period.
 \item \textit{Area under curve (AUC).} We compute the area under the curve of the $\beta_3$ DiD coefficient to capture the total effect size over the observation period.
\end{enumerate}
}

\rev{
In Table \ref{tab:robust}, we give the values of our summary statistics for our main results for questions, answers, and tag groups that we show in Figures \ref{fig:did_qa}, \ref{fig:did_tags}, \ref{fig:did_tags_loc}. and \ref{fig:did_tags_medium}. In addition, we conduct four different robustness checks frequently adopted in observational studies \cite{meyer1995, rosenbaum2020} and, in particular, in DiD designs \cite{lechner2011, roth2023} including placebo event timing \cite{abadie2015, eggers2024}, controlling for co-occurring and potentially confounding events \cite{angrist2009, roth2023}, bootstrap confidence intervals \cite{bertrand2004, cameron2008, efron1994}, and permutation tests \cite{bertrand2004, buchmueller2011, mackinnon2020}. In the following subsections, we contextualize the metrics listed in Table~\ref{tab:robust} and relate them to the robustness of our analyses and results.}

\begin{table}[t]
\centering
\caption{\rev{\textit{Results summary statistics.} For all of our regression results, we show the longest positive effect streak (LPES) as the percentage of the duration of the post-ChatGPT observation period ($151$ days), the maximum of the mean effect size over $151$ one-month rolling windows (MRM), and the total effect over the observation period as the area under the effect size curve (AUC) per day. We normalize the total AUC by dividing with $151$ days for comparison between robustness checks. We assess the robustness of these results by comparing the uncertainty in summary statistics between robustness experiments.
}}
\rev{
\begin{tabular}{l|r|r|r|r}\hline
\textbf{Dataset} & \textbf{Outcome} & \textbf{LPES ($\%$)} & \textbf{MRM} & \textbf{AUC (/day)}\\\hline\hline
Questions & Question Length (Log) & $100\%$ & $0.075$ & $0.063$\\\hline
Answers & Answer Length (Log) & $100\%$ & $0.041$ & $0.029$\\\hline
Web & Question Length (Log) & $100\%$ & $0.115$ & $0.073$\\
Web & Question LOC (Log) & $82.78\%$ & $0.088$ & $0.050$\\
Web & Question Medium Diff. & $48.34\%$ & $0.063$ & $0.036$\\\hline
Python & Question Length (Log) & $100\%$ & $0.138$ & $0.105$\\
Python & Question LOC (Log) & $100\%$ & $0.177$ & $ 0.097$\\
Python & Question Medium Diff. & $83.44\%$ & $0.035$ & $0.027$\\\hline
Java & Question Length (Log) & $64.90\%$ & $0.104$ & $0.072$\\
Java & Question LOC (Log) & $41.06\%$ & $0.083$ & $0.040$\\
Java & Question Medium Diff. & $41.06\%$ & $0.093$ & $0.040$\\\hline
\end{tabular}
}
\label{tab:robust}
\end{table}

\subsubsection{\rev{Placebo launch date}}
\rev{
In this test, we take several placebo ChatGPT launch dates and repeat our DiD regression computation. In particular, we take three placebo event dates before (October 31, September 30 and August 31, 2022) and three dates (December 31, 2022, as well as January 31 and February 28, 2023) after the real ChatGPT launch date. For the placebo events before the real event, we use all the available data in the before period (five, four, and three months, respectively) and six months of data after the placebo event. In case of placebo events after the real event, we use all the available data in the after period (accounting again to five, four, and three months respectively) and six months of data before the placebo event. We show the results of the placebo test for the \textit{python} questions in Table \ref{tab:placebo}. The results for all dataset selections including questions, answers, \textit{web} and \textit{java} questions, show similar trends and we omit them from the table due to presentation reasons.
}

\begin{table}[t]
\centering
\caption{\rev{\textit{Placebo tests.} We show the longest positive effect streak (LPES), the maximum mean effect size (MRM), and the area under the curve (AUC), computed on the regression results with placebo events placed $\pm(1-3)$ months from the real ChatGPT launch. For presentation purposes, we show only results for \textit{python} questions but we observe similar results for all other dataset selections. With $\uparrow$, $\downarrow$, and $\leftrightarrow$ we denote that a given summary statistics increases, decreases, or remains equal to the regression results with the real event date. We observe shorter effect streaks, lower effect size peaks, and lower total effect sizes, indicating that our main results are robust with respect to alternative past or future event dates.}}
\rev{
\begin{tabular}{l|r|r|r|r}\hline
\textbf{Outcome} & \textbf{Placebo Event Date} & \textbf{LPES ($\%$)} & \textbf{MRM} & \textbf{AUC (/day)}\\\hline\hline
Question Length (Log) & August 31, 2022 & $55.92\%$ $\downarrow$ & $0.096$ $\downarrow$ & $0.043$ $\downarrow$\\
Question Length (Log) & September 30, 2022 & $75.65\%$ $\downarrow$ & $0.103$ $\downarrow$ & $0.063$ $\downarrow$\\
Question Length (Log) & October 31, 2022 & $96.05\%$ $\downarrow$ & $0.130$ $\downarrow$ & $0.089$ $\downarrow$\\\hline
Question Length (Log) & December 31, 2022 & $100\%$ $\leftrightarrow$ & $0.132$ $\downarrow$ & $0.107$ $\uparrow$\\
Question Length (Log) & January 31, 2023 & $100\%$ $\leftrightarrow$ & $0.119$ $\downarrow$ & $0.102$ $\downarrow$\\
Question Length (Log) & February 28, 2023 & $100\%$ $\leftrightarrow$ & $0.100$ $\downarrow$ & $0.094$ $\downarrow$\\\hline
Question LOC (Log) & August 31, 2022 & $54.61\%$ $\downarrow$ & $0.066$ $\downarrow$ & $0.032$ $\downarrow$\\
Question LOC (Log) & September 30, 2022 & $73.03\%$ $\downarrow$ & $0.081$ $\downarrow$ & $0.044$ $\downarrow$\\
Question LOC (Log) & October 31, 2022 & $93.42\%$ $\downarrow$ & $0.136$ $\downarrow$ & $0.068$ $\downarrow$\\\hline
Question LOC (Log) & December 31, 2022 & $100\%$ $\leftrightarrow$ & $0.175$ $\downarrow$ & $0.110$ $\uparrow$\\
Question LOC (Log) & January 31, 2023 & $100\%$ $\leftrightarrow$ & $0.167$ $\downarrow$ & $0.119$ $\uparrow$\\
Question LOC (Log) & February 28, 2023 & $100\%$ $\leftrightarrow$ & $0.152$ $\downarrow$ & $0.128$ $\uparrow$\\\hline
Question Medium Diff. & August 31, 2022 & $23.68\%$ $\downarrow$ & $0.028$ $\downarrow$ & $0.010$ $\downarrow$\\
Question Medium Diff. & September 30, 2022 & $42.76\%$ $\downarrow$ & $0.028$ $\downarrow$ & $0.011$ $\downarrow$\\
Question Medium Diff. & October 31, 2022 & $63.82\%$ $\downarrow$ & $0.031$ $\downarrow$ & $0.022$ $\downarrow$\\\hline
Question Medium Diff. & December 31, 2022 & $100\%$ $\uparrow$ $$& $0.034$ $\downarrow$ & $0.029$ $\uparrow$\\
Question Medium Diff. & January 31, 2023 & $53.93\%$ $\downarrow$ & $0.026$ $\downarrow$ & $0.023$ $\downarrow$\\
Question Medium Diff. & February 28, 2023 & $75.41\%$ $\downarrow$ & $0.024$ $\downarrow$ & $0.022$ $\downarrow$\\\hline
\end{tabular}
}
\label{tab:placebo}
\end{table}

\rev{
In case of placebo event dates before the real ChatGPT launch, we observe shorter effect streaks, as for most of the outcome variables, the effect is close to zero and non-significant in the period before the real event. On the other hand, in case of placebo events after the real event, we still observe long and persistent effect runs but with substantially lower maximum and total effect sizes. Altogether, we observe no sustained effects at the placebo events (particularly for the placebo events before the real event), no persistent effect streaks comparable to the true event (again for the placebo events before the real event), and smaller peaks and total magnitudes of the effect (for both the placebo events before and after the real event). Hence, the placebo events do not result in effects comparable to the real ChatGPT launch, providing further evidence on the impact of ChatGPT on Stack Overflow posts.
}

\subsubsection{\rev{Confounding events}}
\rev{
Other platform-related events as well as external factors may potentially confound our results. For example, Stack Overflow platform is regularly updated to include new features or improve the existing features. Such changes may have impact on the user engagement, posting behavior, or post content \cite{anderson2013, kusmierczyk2018, santos2020}. Recent platform updates on Stack Exchange and in particular Stack Overflow are typically published directly on the platform.\footnote{\url{https://meta.stackexchange.com/questions/59445/recent-feature-changes-to-stack-exchange?answertab=createddesc\#tab-top}} Moreover, external events such as new versions of the programming languages, updates to the standard programming libraries or major software packages, or even publishing of new application frameworks may initiate new discussions on the platform and impact the posting intensity, the question and answer content, or the difficulty of the questions. Several sources related to the major programming languages or software packages typically provide detailed histories of such updates.
Last but not least, changes in the service terms or updates to the ChatGPT models may also have confounding effects on our analysis. For example, during our observation period several changes to subscription options and to the ChatGPT models themselves took place.
}

\rev{
To account for these factors, we select seven major events related to (i) platform, (ii) language, and (iii) ChatGPT updates that took place in our observation period. These events include:
\begin{enumerate}[(i)]
\item \textit{Platform changes.} We include two major platform changes that may have impacted the adoption of ChatGPT on Stack Overflow including the ban of ChatGPT answers (December 5, 2022) on the platform and the plagiarism flag (March 22, 2023).
\item \textit{Programming language updates.} We include the major version updates for \textit{python} (October 24, 2022), \textit{javascript} (June 24, 2022), and \textit{java} (September 20, 2022).
\item \textit{ChatGPT updates.} We include the start of the ChatGPT Plus subscription service (February 9, 2023) and the ChatGPT model update to version 4 (March 14, 2023).
\end{enumerate}
While we acknowledge that other major events also happened in this period, with our selection we try to strike a balance between plausibility and potential overfitting or collinearity of possibly overlapping events. We complement our selection of events by a specific regression setup. In particular, for each selected event we introduce a binary variable that we set to one at the event date as well as seven days before the event to account for announcement and anticipation effects and seven days after the event to account for news cycle effects, and to zero otherwise. We then extend our DiD model (Eq. \ref{eq:did}) by controlling for all these additional event variables.
}

\rev{
In Table \ref{tab:events} we show the summary statistics for all combination of outcome variables and dataset selections that we analyze in our papers. Apart from some minimal insubstantial differences to the results of our main DiD regression (Eq \ref{eq:did}), our results remain stable, suggesting the validity of our findings.
}

\begin{table}[t]
\centering
\caption{\rev{\textit{Confounding events.} We show our summary statistics while controlling for events potentially confounding our results. We observe only small $\pm$ differences to the results of our main DiD model (cf. Table \ref{tab:robust}), further confirming robustness of our findings.}}
\rev{
\begin{tabular}{l|r|r|r|r}\hline
\textbf{Dataset} & \textbf{Outcome} & \textbf{LPES ($\%$)} & \textbf{MRM} & \textbf{AUC (/day)}\\\hline\hline
Questions & Question Length (Log) & $100\%$ & $0.073$ & $0.062$\\\hline
Answers & Answer Length (Log) & $100\%$ & $0.040$ & $0.030$\\\hline
Web & Question Length (Log) & $91.39\%$ & $0.113$ & $0.068$\\
Web & Question LOC (Log) & $82.78\%$ & $0.083$ & $0.046$\\
Web & Question Medium Diff. & $35.09\%$ & $0.062$ & $0.035$\\\hline
Python & Question Length (Log) & $100\%$ & $0.142$ & $0.107$\\
Python & Question LOC (Log) & $100\%$ & $0.178$ & $ 0.102$\\
Python & Question Medium Diff. & $82.78\%$ & $0.049$ & $0.031$\\\hline
Java & Question Length (Log) & $41.06\%$ & $0.109$ & $0.072$\\
Java & Question LOC (Log) & $31.78\%$ & $0.087$ & $0.040$\\
Java & Question Medium Diff. & $32.45\%$ & $0.095$ & $0.037$\\\hline
\end{tabular}
}
\label{tab:events}
\end{table}

\subsubsection{\rev{Bootstrap confidence intervals}}
\rev{
We apply block bootstrap, which is a variant of the standard bootstrap resampling procedure that maintains the autocorrelation structure by keeping the blocks of data that belong to the same temporal interval together. In particular, as we find moderate weekly autocorrelation cycles, we randomly sample a starting date and then keep subsequent seven days of data in our bootstrap sample. To draw a complete bootstrap sample, we repeatedly draw starting dates uniformly at random until we have the same amount of days in our sample as in the original data. To compute bootstrap confidence intervals on our summary statistics we draw $1,000$ bootstrap samples, and refit the DiD models on all combinations of data selections and outcome variables.
}

\rev{
We present the result of block bootstrap in Table \ref{tab:boot}. Bootstrap confidence intervals are consistent with our previous findings, indicating robustness of our main DiD results on the effect of ChatGPT on Stack Overflow. 
}

\begin{table}[t]
\centering
\caption{\rev{\textit{Bootstrap confidence intervals.} For all of our regression results, we show the bootstrap confidence intervals for the longest positive effect streak (LPES), the maximum of the mean effect size (MRM), and the total effect as the area under the effect size curve (AUC). Results with bootstrap samples provide further evidence on the robustness of our main DiD results.}}
\rev{
\begin{tabular}{l|r|r|r|r}\hline
\textbf{Dataset} & \textbf{Outcome} & \textbf{LPES ($\%$)} & \textbf{MRM} & \textbf{AUC (/day)}\\\hline\hline
Questions & Question Length (Log) & $(44.35\%, 100\%)$ & $(0.055, 0.095)$ & $(0.038, 0.072)$\\\hline
Answers & Answer Length (Log) & ($19.87\%, 100\%)$ & $(0.018, 0.054)$ & $(0.011, 0.034)$\\\hline
Web & Question Length (Log) & $(23.82\%, 100\%)$ & $(0.054, 0.151)$ & $(0.032, 0.089)$\\
Web & Question LOC (Log) & $(8.61\%. 88.74\%)$ & $(0.035, 0.118)$ & $(0.014, 0.067)$\\
Web & Question Medium Diff. & $(9.27\%, 72.28\%)$ & $(0.029, 0.086)$ & $(0.011, 0.051)$\\\hline
Python & Question Length (Log) & $(31.77\%, 100\%)$ & $(0.068, 0.172)$ & $(0.044, 0.120)$\\
Python & Question LOC (Log) & $(22.5\%, 100\%)$ & $(0.067, 0.209)$ & $(0.037, 0.118)$\\
Python & Question Medium Diff. & $(0\%, 54.97\%)$ & $(0.012, 0.068)$ & $(-0.003, 0.039)$\\\hline
Java & Question Length (Log) & $(13.24\%, 100\%)$ & $(0.068, 0.179)$ & $(0.024, 0.115)$\\
Java & Question LOC (Log) & $(5.96\%, 64.92\%)$ & $(0.039, 0.158)$ & $(0.005, 0.082)$\\
Java & Question Medium Diff. & $(3.97\%, 52.98\%)$ & $(0.029, 0.130)$ & $(-0.002, 0.067)$\\\hline
\end{tabular}
}
\label{tab:boot}
\end{table}

\subsubsection{\rev{Permutation tests}}
\rev{
Finally, we perform permutation tests for all combinations of outcome variables and dataset selections. In particular, we test whether we can reject the null hypothesis of ChatGPT launch having no effect on the posting behavior and content on Stack Overflow. Particularly, if there is no treatment effect and we only observe differences between the treatment and control years due to seasonality, general trends, or some alternative external or internal factors, then the treatment label ($T$) can be randomly reassigned between the posts without significantly changing the overall result. Hence, to compute the test we count how many times the value of a given statistics in permuted data is more extreme than the observed value of that statistics in the real data. We then compute the p-value of rejecting the null hypothesis by dividing this count of extreme values with the total number of permutations. To preserve the temporal structure and, in particular, the weekly and daily seasonal patterns that we observe with the autocorrelation function, we start by grouping the data by the treatment, the week of the year, and the day of the week and then swapping the outcome variables uniformly at random between the control and treatment year, hence, retaining the week of the year and the day of the week for each post in permuted data. Following this random relabeling approach, we compute $1,000$ random permutations, and compute the p-values for our summary statistics.
}

\rev{
Across all data selections including questions, answers, \textit{python}, \textit{web}, and \textit{java} questions and all outcome variables, we consistently observe that the effect completely disappears as measured by the longest positive run, maximum effect size and the total effect size. In particular, a huge majority of p-values (25 out of 33 tests) equals to zero (none of the $25 \times 1000$ permutations resulted in the statistics values more extreme than the observed statistics values), while in the remaining cases the largest permutation p-value is $0.009$ in the case of \textit{java} questions and the longest positive run statistics for the logarithm of the code length and the probability of the medium difficulty questions. Hence, these results provide further evidence in favor of the substantial effect of ChatGPT on Stack Overflow.
}

\subsection{\rev{Effect heterogeneity}}

\rev{To gain insight into the underlying mechanisms of the behavioral changes on Stack Overflow, we conduct an additional stratified DiD analysis to examine heterogeneous ChatGPT effects across user subpopulations defined by propensity scores \cite{imai2022, rosenbaum2020}. This approach allows us to assess whether ChatGPT had differential effects across user segments with varying baseline characteristics. In particular, using only data from the period before ChatGPT launch (or before November 30, 2021 for the control year), we extract total user activity, total number of answers and comments that their questions received, as well as total number of views and total score of their questions as user features. With this approach to selection of data, we effectively require that users have at least one post in the period before treatment. In addition, from the selected users we also exclude all the users that did not have any activity in the period after the treatment start. This substantially reduces the total number of users for further processing. For example, for questions the total number of users is reduced from $1,102,385$ to $680,492$, and we observe similar reductions of around $40\%$ in all other cases as well.
Next, we transform user features by taking logarithms (except for the total score) to account for the skew in their distributions. We then fit a logistic regression with the treatment variable $T$ as the outcome. Lastly, for each user, we take the predicted probability from the logistic model as the propensity score of being treated and assign this score to all posts made by that user.} 

\rev{
Using propensity scores, we proceed by computing post weights as average treatment effect of the treated (ATT) weights. With this weighing scheme treated users are weighted with $1$ and control users with $s_i / (1 - s_i)$, where $s_i$ is the propensity score of user $i$. We then first use these ATT weights to check for the common support and balance. In particular, to check for the balance we estimate standardized mean differences (SMD) between the treated and the control group for all user features, and to check the common support we visually investigate the distributions of the propensity score between two groups. 
We perform these checks on all users and separately on selection of user subpopulations. Hence, we divide users into three segments using the quartiles of the propensity score distribution: (i) first quartile (Q1) of the propensity score distribution with the users having a low probability of being treated, (ii) second and third quartiles (Q2-Q3) of the distribution, which keeps users where the treatment probability is almost coin flip (e.g., for questions the range of propensity scores in this user group is $0.39$ to $0.54$), and (iii) fourth quartile (Q4) of the distribution with the users having the highest probability of being treated.}

\rev{
Before proceeding with the propensity score checks, we first characterize user segments by inspecting the distributions of their features. We find that, in all cases, the Q4 user segment tends to include users with lower overall activity, lower numbers of received answers and comments, lower number of question views, and lower scores. As Q4 users have the highest probability of being treated, this group matches treated users with control users that are most similar to the treated users and, hence, captures the general downward trend in user activity that we observe on Stack Overflow in our data (cf. Fig. \ref{fig:q}). On contrary, Q1 segment tends to include users with substantially higher values of all user features. As Q1 includes users with the lowest probability of being treated, this segment also captures the overall negative trend on Stack Overflow. In particular, users with substantial amount of activity have low probability of being active in the treatment year. Lastly, Q2-Q3 segment, including users with propensity scores close to $0.5$, tends to match users with feature values in the mid-range, matching regular users from both years to each other.
}

\rev{
Continuing with the balance and support checks, we obtain an excellent balance with SMD values significantly smaller than $0.01$ in almost all cases. In rare cases, where we have fewer data points after our user selection and segmentation procedure such as Q4 users for \textit{java} or \textit{python} questions, we obtain the largest SMD of around $0.25$ for an individual feature. Moreover, our analysis of the common support where we check the overlap of propensity score distributions after ATT reweighting, reveals no substantial mismatch between the propensity score distributions between the treated and the control group in neither of the cases. Hence, these checks indicate a well balanced user segmentation with an excellent common support for our particular selection of user features.
}

\rev{
Next, we combine ATT weights with DiD regressions. These weighting scheme weights the control posts that are most similar to the treated posts with higher weights, effectively making the control and the treated groups more comparable. We start with an additional robustness check by fitting ATT weighted DiD regressions on all combinations of the outcome variables and the data selections and with all extracted users (without segmenting them into Q1, Q2-Q3, and Q4 yet). We obtain even more pronounced effects than with the original data. In particular, all our summary statistics substantially increase, hence, providing further evidence for the stability and robustness of our findings. 
}

\rev{
Finally, we fit the DiD regression for all data selections and the outcome variables, separately for each user segment. For questions and answers, we find more pronounced effects for Q1 and Q2-Q3 segments, while the effect is not robustly present in the case of Q4 segments. The results for the individual programming languages further amplify this finding. In particular, for all three programming languages the effect is substantial and statistically undistinguishable from our main results in the Q2-Q3 user segment, as measured by our summary statistics. While for \textit{python} and \textit{web} in Q1 segment, the effect is still present but slightly weaker than in the complete dataset, in case of a smaller \textit{java} dataset, the effect disappears for Q1 users. In all three cases, the effect is not robustly present in Q4 segment. From those results, we draw the following conclusions. We observe a substantial effect heterogeneity across user segments defined by their overall activity and the community attention that they receive in the form of answers, comments, views, and the overall question score. In particular, the effect is not present in users with low activity levels that gain less attention (Q4). We hypothesize that questions gaining less attention from less experienced users potentially deal with well-known, already resolved or even simpler questions. Hence, these questions do not substantially increase in length or difficulty between the treated and control year. Also, the effect is somewhat weaker and less robust for the most active users and their questions that receive larger amounts of attention from the community (Q1), in particular for smaller subpopulations such as \textit{java}. We hypothesize that the most active and therefore more experienced users post questions, which are longer and potentially deal with more advanced problems. Such problems, at least immediately after the introduction of ChatGPT, were not easily solved by ChatGPT, and hence, they were less affected by its introduction. Lastly, the effect is at the strongest in the mid-range user segment Q2-Q3, in which users are more regularly active and obtain more attention than users in Q4. We hypothesize that in this segment, the users post more advanced programming questions than in segment Q4 but potentially less advanced than in segment Q1. We argue that ChatGPT was probably able to provide satisfactory answers to a large portion of these questions, and, hence, we observe a strong drift towards longer and more difficult questions in this user segment.
}

\begin{figure}[t]
	\centering
	\subfloat[\rev{Q1 user segment}]{\includegraphics[width=0.33\textwidth]{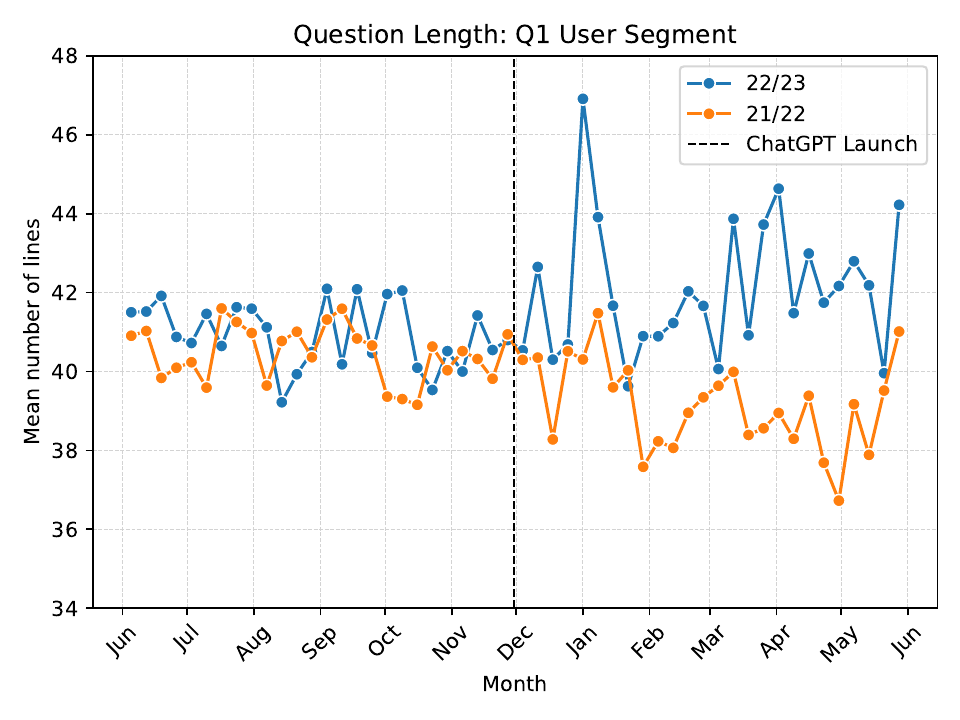}\label{fig:q1}}
    \subfloat[\rev{Q2-Q3 user segment}]{\includegraphics[width=0.33\textwidth]{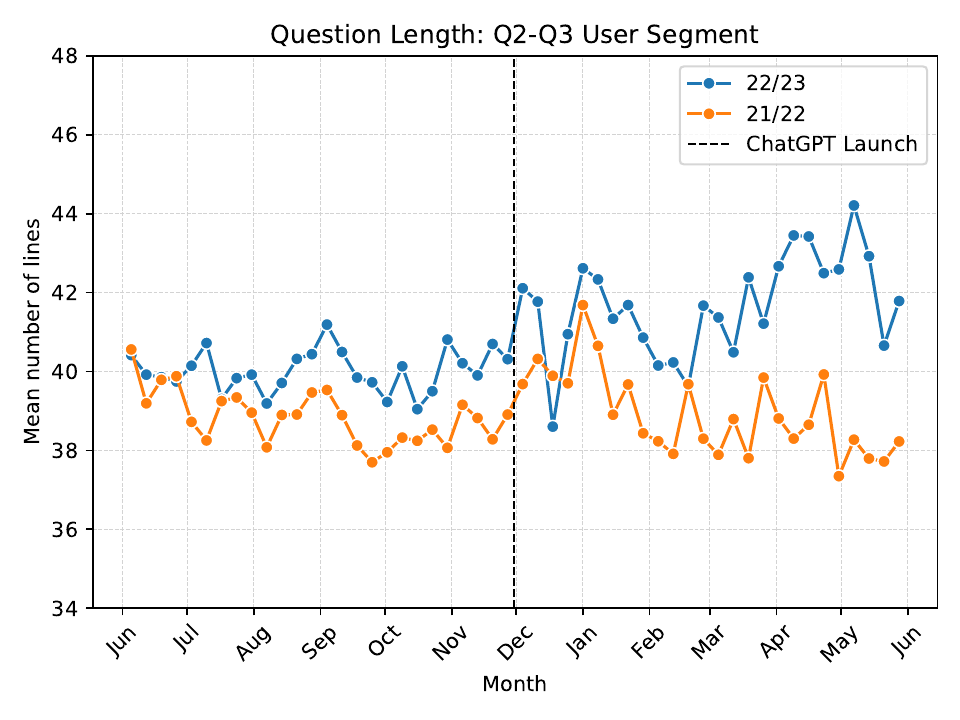}\label{fig:q23}}
    \subfloat[\rev{Q4 user segment}]{\includegraphics[width=0.33\textwidth]{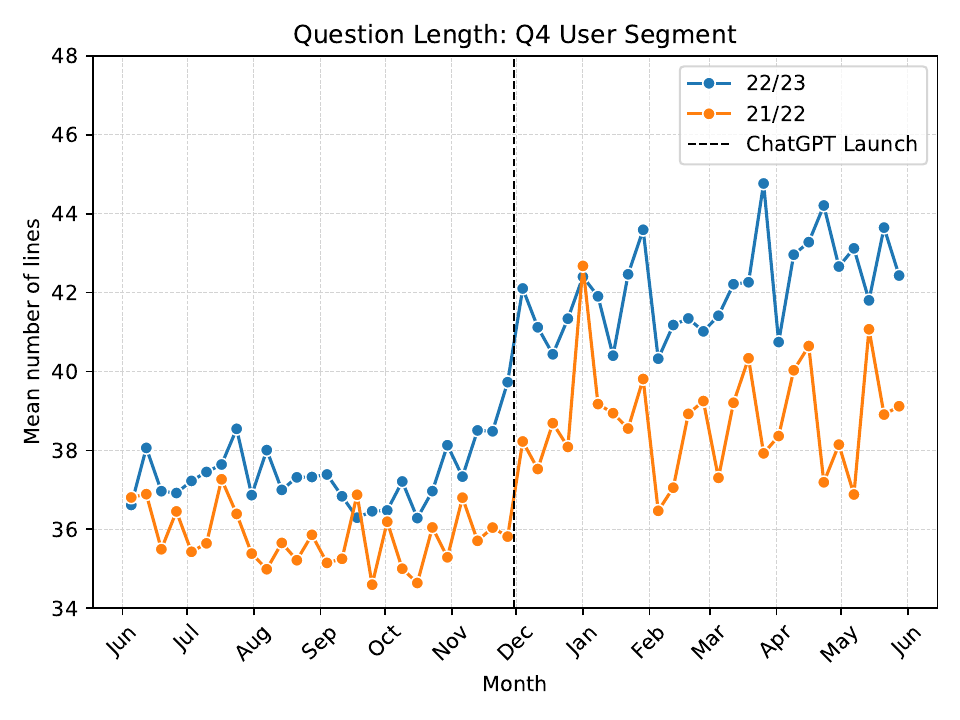}\label{fig:q4}}
	\caption{\rev{\textbf{User segmentation with propensity scores.} We compare the question length in user segments identified with the quartiles of their propensity score distribution. In (a) we show Q1 containing users that are least likely to receive treatment. This user segment contains most active users gaining a lot of attention. In (b) we show Q2-Q3 user segment where the probability of being treated is almost a coin flip. This group of users contains regular users with a substantial attention. Lastly, we show Q4 segment with the least active users that gain less attention in (c). Neither already longer questions from experienced Q1 users (a) nor shorter questions from less active users (c) exhibit a strong differential increase in length in the treated year as compared to the control year. The most affected are mid-range questions from the regular users in Q2-Q3 (b).}}
	\label{fig:segments}
\end{figure}

\rev{
We substantiate our explanations and hypotheses with an analysis of the outcome variables per user segment. In particular, we look at the question length and difficulty and compare their temporal developments for each user segment. In Figure \ref{fig:segments}, we show the weekly question length means. We observe that the questions are longest in Q1 segment and then gradually become shorter in Q2-Q3 and Q4. We find that the questions of the users in Q4 segment in both the treated as well as the control year gradually increase in length with time. This potentially suggests that the users write longer questions as they gain more experience with the community (recollect that Q4 users are the least active of all user segments) or as they learn how to phrase the questions to gain more attention from the community. However, the increase in question length takes place in both years, suggesting a systematic posting behavior drift of less active users with their tenure (see Fig. \ref{fig:q4}). On the other hand, in the group of Q1 users (most active users before the treatment start) we find a quite stable question length in both groups over time, particularly in the treated year (see Fig. \ref{fig:q1}), resulting in a less pronounced effect in this user segment. Note that there is a slight negative trend of the question length in the control group suggesting that in this user segment ChatGPT potentially reversed this downward trend. Finally, in the group of regular users (Q2-Q3), we observe a positive trend in the question length over time for the treated group as compared to the control group where the question length remains quite stable (see Fig. \ref{fig:q23}). We observe similar trends in other relevant outcome variables such as question difficulty as well as in subgroups related to individual programming languages, which we do not show here for space reasons. Hence, we conclude that the behavioral change on Stack Overflow is the strongest in the group of regular users and that both, the most active and experienced, as well as, the least active and experienced users, are less affected by the ChatGPT launch. 
}


\subsection{Question content drift}

To gain additional insight in the nature of the change in the difficulty of the programming questions on Stack Overflow, we extend our analysis to the content of the questions. Specifically, we use BERTopic \citep{grootendorst2022} to cluster the CodeT5 embeddings that we computed earlier. To gain insight in the question content, we also extract the most representative words from each cluster after removing common stopwords, as well as highly frequent words. More specifically, starting with the data from the 22/23 period, we divide that data into pre-ChatGPT and post-ChatGPT periods. Then, from each of those periods we  extract eight topics to analyze the topical shifts in questions before and after the ChatGPT launch. We repeat the topic analysis for each of our tag groups. Finally, we evaluate the quality of our topic extraction by computing the silhouette coefficient for each of the 16 topic models. We obtain the silhouette coefficients ranging from $0.299$ (smallest value obtained for \textit{C\#} tag post-ChatGPT) to $0.401$ (largest value obtained for \textit{web} tag group post-ChatGPT) indicating fairly well divided topics. 

\begin{figure}[t!]
	\centering
    \subfloat[Python: topic analysis pre-ChatGPT]{\includegraphics[width=0.8\textwidth]{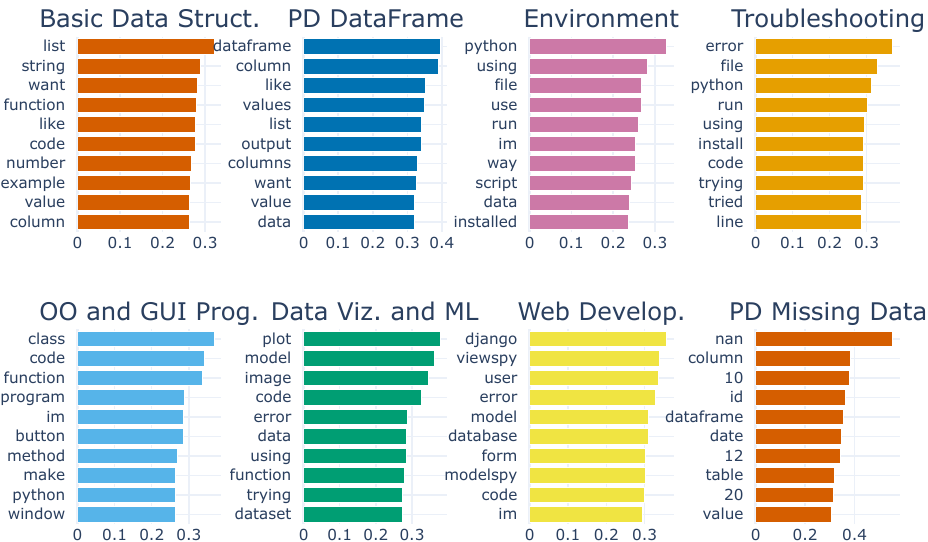}\label{fig:python_before}}\\
    \subfloat[Python: topic analysis post-ChatGPT]{\includegraphics[width=0.8\textwidth]{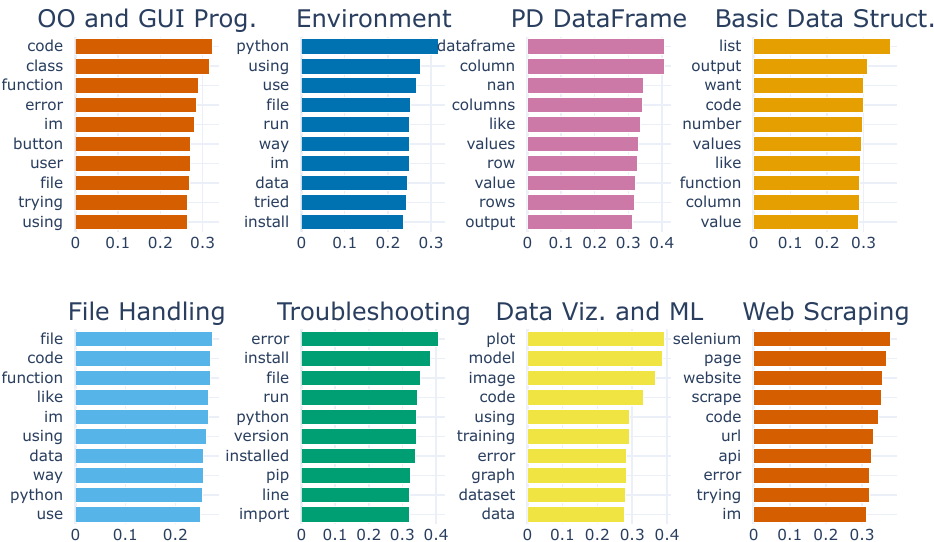}\label{fig:python_after}}
	\caption{\textbf{Content drift on Stack Overflow.} We extract eight topics from \textit{python} questions before the ChatGPT launch and another eight topics after the launch. While the topics remain relatively stable, their relative sizes change. Before ChatGPT launch, the largest category of questions was related to \textit{Basic Data Structures}, which are typically related to easier programming tasks. After the introduction of ChatGPT, the largest topical category is related to more difficult questions on \textit{Object-oriented and GUI programming}.}
	\label{fig:topics}
\end{figure}

In Figure \ref{fig:topics} we show the extracted topics sorted by the number of questions that they include, together with their most representative words for \textit{python} questions (we omit remaining topic models due to space constraints). We create the topic labels by prompting ChatGPT with the most representative words of the corresponding topic. We find that the \textit{python} topics are fairly stable across the periods with slight changes in the smaller topics (cf. shift from \textit{Web development} to \textit{Web scraping} and shift from \textit{Pandas missing data} to \textit{File handling} in Figure \ref{fig:topics}). However, we observe a shift in the number of questions per topic, as well as slight distributional changes regarding the most representative words. Most prominent changes are related to topics \textit{Object-oriented and GUI programming} and \textit{Basic Data Structures}. While \textit{Basic Data Structures} drops from the largest topic pre-ChatGPT (17,75\% of questions) to fourth largest post-ChatGPT (12,98\% of questions), \textit{Object-oriented and GUI programming}, climbs from the fifth largest topic pre-ChatGPT (13,95\%) to the largest topic post-ChatGPT (20,10\%). In programming, basic programming concepts such as variables, loops, or basic data structures, or functions are considered a subset of more sophisticated object-oriented programming concepts (cf. ACM Computer Science Curricula \citep{kumar2024}). As such, object-oriented concepts are typically taught later in computer science curricula than the introductory programming concepts. We observe similar shifts towards more sophisticated topics in other tag groups, but they are typically tag specific. For example, in \textit{java} we observe an increase in questions on the Spring Web development framework, while \textit{C\#} also experiences an increase in questions about object-oriented programming albeit a weaker one than \textit{python}.

\rev{
As a robustness analysis, we repeat the topic extraction for \textit{python} questions with six and ten topics. While the results slightly vary as some topics (i) are merged into larger ones, (ii) are split into sub-topics, (iii) appear as new topics, or (iv) are represented with differing words, we obtain quantitatively and qualitatively comparable results to the extraction with eight topics. In particular, when extracting six topics we observe a slightly lower silhouette coefficient ($0.297$ as compared to $0.342$ with eight topics) in pre-ChatGPT period and a slightly higher silhouette coefficient ($0.340$ as compared to $0.336$ with eight topics) in post-ChatGPT period. In this extraction, a new topic \textit{Functions} that includes several keywords previously present in the \textit{Object-oriented programming} topic becomes the largest topic in post-ChatGPT period, while the \textit{Object-oriented programming} topic is the second largest topic in both pre-ChatGPT and post-ChatGPT periods. The largest topic before the introduction of ChatGPT \textit{Troubleshooting} (dealing mostly with installation problems and import statements) is split into two smallest topics post-ChatGPT. In the case of ten topics, we observe higher silhouette coefficients in both periods ($0.367$ pre-ChatGPT and $0.354$ post-ChatGPT) but a similar content drift as with eight topics. In particular, the largest pre-ChatGPT topic \textit{Basic Data Structures} drops to the fourth position in the post-ChatGPT period, while the fourth largest pre-ChatGPT \textit{Object-oriented programming} topic becomes the second largest topic in the post-ChatGPT period. The \textit{Environment} topic becomes the largest topic post-ChatGPT (the second largest in the post-ChatGPT period with eight topics, cf. Fig \ref{fig:python_after}).
}

\section{Discussion}

\subsection{\rev{Summary of results}}
With our study, we confirm previous findings of an accelerated decline of questions and answers on Stack Overflow post-ChatGPT \citep{burtch2024, delrio2024}. However, we find that the length of the questions and answers, as well as, the length of code examples in questions substantially increases in both, short-term (e.g., a few weeks after the launch), as well as long-term (up to six months after the launch) periods. Moreover, we find that question difficulty after the ChatGPT introduction significantly increases and shifts from easy questions towards medium difficulty questions as estimated by our XGBoost classifier. \rev{We also observe differential effects of ChatGPT in user segments on Stack Overflow. In particular, regular users exhibit the strongest response, while experienced and sporadic users show a weaker response or no response at all, respectively.} Additionally, we observe a content drift across thematic categories towards questions related to more sophisticated concepts such as object-oriented programming. \rev{Lastly, we conduct a range of robustness checks such as placebo or permutation tests. Overall, the evidence from these robustness experiments points overwhelmingly to a substantive and stable effect of ChatGPT on Stack OVerflow.}

\subsection{\rev{Ask ChatGPT for simpler and the crowd for more difficult questions}}
We suggest that Stack Overflow community experiences an ongoing fundamental behavioral shift related to the introduction of ChatGPT. In particular, our results indicate that users, at an accelerating rate, turn to the Stack Overflow crowd in case of more difficult questions after the ChatGPT launch. \rev{We hypothesize that users increasingly ask ChatGPT for answers to simpler or well-known programming questions as ChatGPT provides immediate and reasonably good answers in majority of such cases \citep{austin2021, chen2021, luo2023}. However, as we do not directly observe how users work with ChatGPT, we leave the investigation of this hypothesis to future research.}
We corroborate our large-scale quantitative results and our conclusions with the results of several recent studies related to ChatGPT, Stack Overflow, and programming tasks and questions. For example, Kabir et al. \citep{kabir2024} analyze ChatGPT answers to programming questions from Stack Overflow and find that more than a half of these answers contain incorrect, redundant, or irrelevant information due to its incapability to understand the larger context of the question. Often, code errors produced by ChatGPT are because of wrong logic, wrong reasoning, wrong API calls, or wrong function calls, which may indicate that ChatGPT is struggling when logical or algorithmic reasoning or coordination of external APIs or function calls is required, all defining features of more advanced programming questions. At the same time, ChatGPT makes fewer syntax errors and generally makes fewer errors when answering more popular or older Stack Overflow questions, suggesting that it handles simpler or well-established question and answers better. In another study by Widjojo and Treude \citep{widjojo2023}, the authors find that LLMs can compete and outperform Stack Overflow answers on questions related to compiler errors. However, a higher quality in LLM responses to coding questions and tasks is typically related to experience in creating efficient prompts. In particular, the authors find that carefully designed prompts substantially increase the quality of LLM answers, indicating that a certain level of experience in creating prompts is needed to achieve more precise results. Along similar lines, further studies found evidence for usefulness of ChatGPT, Copilot and other general-purpose LLMs for a variety of programming and programming assistance tasks \citep{peng2023, ross2023}. In particular, the participants of the study by Ross et al. reported the usefulness of an LLM programming assistant for ordinary, simpler, small, and repetitive tasks, as well as for short chunks of code, little coding problems, or quick lookups \citep{ross2023}. Similarly, Chen et al. \citep{chen2021}, as well as Yetistiren et al. \citep{yetistiren2022} found that fine-tuned LLMs can produce a high quality code on easy interview problems for software developers. 

\subsection{\rev{ChatGPT improvements and user experience}}
As we only analyzed the first six months after start of the ChatGPT, shift of users to Stack Overflow for more difficult questions may be potentially due to the initial ChatGPT-related issues, and may cease with improved model versions. For example, some of the early ChatGPT problems were frequently related to deficiencies in analytic thinking and reasoning, problems later addressed by, among other approaches, chain-of-thought prompting \citep{wei2022}. However, difficulties of LLMs in logical and algorithmic reasoning \citep{naveed2025, shojaee2025, valmeekam2022} are still present and combined with their limited domain knowledge (recently increasingly addressed by techniques such as Retrival Augmented Generation \citep{gao2024}), frequent hallucinations \citep{huang2025}, or their scaling limitations \citep{naveed2025}, still remain limiting factors in their adoption for various tasks including problem solving tasks in programming. 

In addition to the limitations of the LLMs in handling more sophisticated problems, the experience in working with LLMs and the ability of users to refine prompts substantially contributes to the quality of answers \citep{widjojo2023}. Combination of those two factors, i.e., intrinsic LLM limitations as well as users requiring time to gain experience, may potentially explain the accelerating trend towards Stack Overflow in case of more complex tasks (cf. Figures \ref{fig:did_tags} and \ref{fig:did_tags_loc} for trend in the question length and the length of code examples, and Figure \ref{fig:did_tags_medium} for trend in question difficulty) as more and more users are able to obtain satisfactory results from ChatGPT on more sophisticated questions with time, practically setting the benchmark for Stack Overflow questions higher.

\subsection{\rev{ChatGPT's disruption of Stack Overflow}}
The original theory of disruptive technology introduced by Christensen \citep{christensen1997} defined a disruptive technology as a technology inferior in the main attributes to the mainstream technology but with innovative features that the mainstream technology neglects. Later definitions of the disruptive technology focused more on the characteristics of the new technology and less on the niche aspects introduced by the original theory. For example, Nagy et al. \citep{nagy2016} define a disruptive innovation as ``An innovation that changes the performance metrics, or consumer expectations, of a market by providing radically new functionality, discontinuous technical standards, or new forms of ownership.'' In that sense, due to LLMs' remarkable performance in various domains \citep{teubner2023} such as education \citep{garcia2023}, medicine \citep{chow2023, heng2023},  research \citep{gao2023}, or software development \citep{liu2023}, ChatGPT and other LLMs may be already considered as a disruptive innovation. 

More specifically, we now compare our study on effects and disruptive aspects of ChatGPT on help-seeking in software development with studies analyzing effects of ChatGPT on other activities in collaborative knowledge work. For example, in traditional knowledge work, studies assessing the worker productivity in consulting \citep{dell2023} found differential effects of artificial intelligence tools on the consultants' productivity, suggesting that ChatGPT had a positive effect only on some of the tasks (e.g., creative product innovation), while using ChatGPT in other tasks (e.g., data and interview analysis) had deteriorating effects. Along those lines, Noy and Zhang \citep{noy2023} found that on writing tasks ChatGPT increased the authors' productivity as well as the quality of the written texts suggesting a positive effect of ChatGPT on this particular knowledge creation task. Potentially most related to our study, a recent study \citep{reeves2024} analyzed page views, visitors, edits, and editors across languages to quantify the effect of ChatGPT  on Wikipedia, another popular collaborative knowledge creation platform. The authors found that across all languages, all metrics increased after the introduction of ChatGPT but the increase was smaller for languages where ChatGPT is available, suggesting a diverse effect dependent on the Wikipedia language edition. Similar to those works, our study gives insight into differential effects of ChatGPT on help-seeking platforms for software developers---while previous studies of ChatGPT and Stack Overflow found decaying number of questions and answers as the consequence of ChatGPT \citep{burtch2024, delrio2024}, we find that the questions, answers, and code examples are longer. Additionally, the question difficulty increases and the question content shifts towards more advanced topics after the ChatGPT start.

As our results for the first six months of ChatGPT suggest, we expect the observed disruption of Stack Overflow due to ChatGPT to continue in future. In particular, we argue that (i) as LLMs mature, (ii) as they are further fine-tuned and developed, (iii) as new achievements in prompt engineering are achieved, and (iv) as users gain more understanding and experience in interacting and working with LLMs, we may experience sustained and ongoing shifts in the user behavior with ChatGPT and on question answering platforms. Specifically, we expect users to rely on LLMs for, not only, a wide range of well-established, simpler problems, but also for an ever larger number of more and more difficult questions. This development will set the bar for Stack Overflow questions higher---as more difficult questions are answered by LLMs in a satisfactory manner, the users will continue to turn towards online help-seeking communities for even harder, more complex, more recent, or logically or algorithmically more challenging questions. Hence, we conclude that while ChatGPT disrupted Stack Overflow platform leading to less questions in total, it also started a transformation of the platform, in particular, a transformation of the content posted on Stack Overflow.

\section{\rev{Threats to validity}}

\rev{Although we carefully designed our DiD models and performed a range of robustness checks, our work is not without limitations and there are several threats to the study's internal and external validity.}

\subsection{\rev{Threats to internal validity}}

As in all causal inference, \rev{particularly in observational studies}, we cannot rule out that unmeasured confounders \rev{(omitted variable bias)} still affect our results. \rev{For example, we can not measure external user characteristics such as their level of knowledge, skills, or motivation for their questions. Such user features evolve over time and are typically correlated with our outcome variables such as the question length or the question difficulty. Further, some of the variables that we use, can be incorrectly measured. For example, we measure the question length, the code length or the question difficulty by carefully applying regular expressions, text parsing, and advanced machine learning models but due to unstructured nature of the data in online social platforms these measurements are not without error. Although in general, DiD designs reduce the threats such as omitted variable bias or mismeasurements \cite{meyer1995}, we still can not completely rule out that there are further confounders or unmeasured trends that may distort our findings. However, with our range of robustness checks, we account for many of such potential threats, and the results of these additional checks indicate that are main findings remain stable and robust.}

\rev{Moreover, the outcome variables may have additional trends than the ones that we account for. In particular, with the analysis of the autocorrelation function, we identified weekly and daily patterns in our data. While we controlled for those trends with our DiD design, other user-specific trends may exist. For example, we find that the user behavior changes as users gain experience or adapt to the community norms with time. With the analysis of the user segments, we made an initial step towards analysis of such trends. In addition, by estimating the trends in the ChatGPT's effect on Stack Overflow with the ``secret weapon'' over a period of six months, we further reduce the effects of such user behavioral changes. However, residual user trends such as users naturally writing longer questions as they gain in knowledge, may still affect our results.}

The control group in \rev{our year-over-year design}, is another limitation of our work as this is only a quasi-control group not obtained via an experimental design. With our control group we make an implicit assumption that there are no significant distributional drifts in the users, their behavior, and the content they post (apart from the drift caused by the ChatGPT intervention), i.e., that the control and treatment groups are fairly comparable to each other in the majority of their defining features. We believe that by taking one year of data for both groups \rev{together with our modeling of the seasonality effects}, we account for all seasonal aspects, as well as that sudden changes in behavior are averaged out over such prolonged periods of time. \rev{In addition, with the propensity score calculation and stronger weighting of the control posts that are more similar to the treated posts, we make an important initial step toward increasing the comparability of the control and the treated year. However, as is common in causal inference, we can not guarantee the absence of additional systemic behavioral changes in the data. Potentially, such changes could be further eliminated by combining our DiD regression setup with additional matching strategies not only at the user level, but also at the level of topics or content.} We see this as an interesting avenue for future research in this area.

Finally, other external events such as new versions of the programming languages, new releases of the standard libraries, or introduction of new programming frameworks may also lead to longer and more difficult questions on Stack Overflow. Moreover, during the first six months after ChatGPT introduction, several versions of the ChatGPT model have been released, potentially distorting our results and conclusions. \rev{Even though we tried to account for some of such external events with our analysis of confounding events, there may be other events confounding our findings, which we did not consider. Also, we only concentrated on the events that are either related to Stack Overflow, ChatGPT, or the programming languages, but additional events such as broader social or economic events may still impact our results.}

\rev{
Finally, as with other similar event studies, the timing of the event is typically not exact. ChatGPT had a beta-testing phase and the anticipation for this introduction was already building up in the period before the actual start. Also, community reactions from such events may be gradual and not immediate, as more and more users start to use ChatGPT with time. In our study, we account for after period sensitivity (e.g., in the exact starting date of ChatGPT) of the effect with the rolling window model. Moreover, with placebo event tests we also try to handle potential sensitivities in the period before the ChatGPT launch. With our design decisions, checks for the parallel trend assumption, and additional placebo tests, we aim to (i) reduce uncertainty in the estimates by analyzing data over prolonged periods of time, and (ii) account for any long-term trends on Stack Overflow by finding evidence for the parallel trend assumption between the treatment and the control group, hence, further reducing potential threats to the validity of our results.
}

\subsection{\rev{Threats to external validity}}

\rev{External validity deals with the question whether our results can be generalized to other user populations, collaborative platforms, or different time. While we believe that ChatGPT as an example of a disruptive technology, has a potential effect on other platforms or user populations, we concentrate in our study solely on Stack Overflow and measure the effect of ChatGPT on this platform only. Other platforms have their own context and set of outcome variables, and these platforms may experience different effects. While some of our methods or the general approach can be useful in studying such additional platforms, observational studies such as ours, need always a careful selection of data, models, and analysis steps that are tailored to the platform and specific research questions.}

\section{Conclusions}

\subsection{\rev{Summary \& implications}}
With our large-scale analysis of two years of Stack Overflow data we found that ChatGPT sustainably changed content creation on Stack Overflow. In particular, users tend to write longer questions with longer code examples and questions of an increased difficulty. As ChatGPT is able to provide quick and satisfactory results to simpler, shorter questions, these new circumstances may be also a manifestation of a deeper user behavioral change---modern artificial intelligence tools are relieving the user work burden, in particular burden of repetitive or elementary work activities in software development. While these new conditions result in higher efficiency, more resources, and more time for work on more sophisticated problems, they also raise the question of the future of online collaborative platforms such as Stack Overflow. With ChatGPT and similar tools around, will such programming help-seeking platforms still be necessary in the future? Our results suggest that at the moment they still are, and that they will remain necessary also in the future. However, these platforms will continue to profoundly change as they transition towards new topics, new and more sophisticated content, or new ways of help-seeking in software development.

\subsection{\rev{Future work}} 
The analysis of such transformative future changes represents an interesting direction for future work. For example, a deeper analysis of the Stack Overflow content can reveal more specific drifts in the question topics including emergence of the new topics such as topics related to the use of LLMs themselves. Also, the analysis of user activities, their searching or tagging behavior may provide insights into the specific cases in which users turn to Stack Overflow instead of ChatGPT and may help improve information organization and retrieval functionality on the platform. Finally, analyzing user social interactions, their voting or commenting behavior may provide further actionable information for the operators of collaborative knowledge creation platforms on how to adjust or add new features to better reflect new actuality induced by modern artificial intelligence tools.

\bibliographystyle{elsarticle-num}
\bibliography{paper}

\end{document}